\documentclass[usegraphicx,usenatbib]{mn2e}
\begin{document}

\title[Quiescent neutron stars in NGC 6440 and Terzan 5]{Investigating Variability of Quiescent Neutron Stars in the Globular Clusters NGC 6440 and Terzan 5}
\author[Walsh, Cackett \& Bernardini]{A. R. Walsh$^{1}$\thanks{arwalsh@wayne.edu}, 
E. M. Cackett$^1$ and
F. Bernardini$^{1, 2,3}$
\\$^1$ Department of Physics \& Astronomy, Wayne State University, 666 W. Hancock St., Detroit, MI 48201, USA
\\$^2$ INAF, Osservatorio Astronomico di Capodimonte, Salita Moiariello 16, I-80131 Napoli, Italy
\\$^3$ New York University, Abu Dhabi, P.O. Box 129188, Abu Dhabi, United Arab Emirates
}

\date{Received ; in original form }
\maketitle

\begin{abstract}
The quiescent spectrum of neutron star low-mass X-ray binaries typically consists of two components -- a thermal component associated with emission from the neutron star surface, and a non-thermal power-law component whose origin is not well understood.  Spectral fitting of neutron star atmosphere models to the thermal component is one of the leading methods for measuring the neutron star radius.  However, it has been known for years that the X-ray spectra of quiescent neutron stars vary between observations.  While most quiescent variability is explained through a variable power-law component, the brightest and best-studied object, Cen X-4, requires a change in the thermal component and such thermal variability could be a problem for measuring neutron star radii.  In this paper, we significantly increase the number of sources whose quiescent spectra have been studied for variability.  We examine 9 potential quiescent neutron stars with luminosities $\le 10^{34}$erg s$^{-1}$ over the course of multiple {\it Chandra} observations of the globular clusters NGC 6440 and Terzan 5 and find no strong evidence for variability in the effective temperature in 7 of the 9 sources.  Two sources show a potential change in temperature, though this depends on the exact model fitted.  CX1 in NGC 6440 is equally well fit by a variable thermal component or a variable power law.  Therefore, the results are inconclusive and we cannot exclude or require thermal variability in that source.  CX5 in NGC 6440 shows a potential change in temperature, though this depends on whether a power-law is included in the spectral fit or not.  This suggests that thermal variability may not be widespread among quiescent neutron stars with luminosities $< 10^{34}$~erg s$^{-1}$, and hence thermal radiation remains a promising means to constraining neutron star radii.
\end{abstract}


\section{Introduction}

Neutron stars are the densest directly observable objects in the universe and therefore allow us to probe the behavior of cold matter at supranuclear densities ($\sim10^{15}$ g cm$^{-3}$) otherwise inaccessible.  Discriminating amongst the several plausible dense matter equations of state \citep[e.g.,][]{lattimer04} requires constraining the key parameters, including the neutron star radius.  Thermal processes throughout the neutron star lifetime create a measurable flux which is related to the effective area (and hence radius) and temperature, as for any blackbody-like object. Thus, studying thermal emission from quiescent neutron stars is important for our understanding of matter at extreme densities \citep[see e.g.][and references therein]{guillot13,heinke14}.

Neutron star X-ray transients are a class of low mass X-ray binaries (LMXBs) in which a neutron star intermittently accretes matter from a stellar companion with mass approximately smaller than that of the Sun.  Neutron stars in binary systems cycle through a process of accretion and a state of quiescence.  Periods of enhanced accretion, called outbursts, are relatively short (typically, weeks to months) in these systems, and a neutron star spends the majority of its time in quiescence (years to decades or longer), allowing for examination of its quiescent spectra over time.   During outbursts, the accretion disk brightens significantly, increasing the X-ray luminosity by many orders of magnitude from approximately $L_x = 10^{31} - 10^{34}$ erg s$^{-1}$ in quiescence to approximately $L_x = 10^{37} - 10^{38}$ erg s$^{-1}$  in outburst \citep[though note the emerging class of very faint transient and persistent sources in the intermediate luminosity range, see, e.g.][]{armaspadilla13a,armaspadilla13b,armaspadilla13c}.  Electron captures in the upper layers of the crust and pycnonuclear fusion in deep crustal layers heats the neutron star during outburst \citep{haensel90}.  This process is usually referred to as deep crustal heating.  When accretion is halted, some heat is conducted to the core of the neutron star while the rest diffuses to the surface and is radiated away.  This X-ray emission is widely thought to account for the thermal component of quiescent X-ray spectra that dominates at energies below $2 - 3$ keV \citep[e.g.,][]{brown98}.  A power-law component sometimes dominates the spectra at higher energies.  Its origin is uncertain, but may be due to residual accretion onto the neutron star magnetosphere or a variety of other mechanisms, for instance pulsar shock emission \citep{campana98}.

During quiescence, thermal emission from the neutron star surface allows for a measurement of the neutron star radius.    
However, one complication is that some quiescent LMXBs are variable on a wide range of timescales (minutes to years), in a manner that cannot be reconciled with the deep crustal cooling scenario.  Consequently, the origin of the variability is unknown, or debated, in most cases \citep{campana97,rutledge02,campana04}.  In particular, variability in the thermal component represents a problem as stochastic variability is not expected in the deep crustal cooling model.  Consequently, it is no longer clear whether the inferred emitting radius is correct.  Thus, there is a need to understand the origin of quiescent variability in order to properly assess neutron star radii measurements using quiescent neutron stars.  Variability in the spectra of X-ray transients in quiescence can most often be ascribed to variation in the power-law component with a few exceptions.  \citet{rutledge01} attributed the variability in Aql X-1 to a change in the effective temperature, while \citet{campana03} modeled the spectra of Aql X-1 with a changing column density and variable power-law component, likely due to pulsar shock emission. More recent studies of Aql X-1 by \citet{cackett11} and \citet{cotizelati14} used larger data sets but were still not conclusive in attributing the variability of the quiescent X-ray spectra to the thermal or power-law component. On the other hand, Cen X-4, a nearby quiescent neutron star with the highest known X-ray flux, showed strong evidence of thermal variability during quiescence  \citep{cackett10, cackett13_cenx4, bernardini13}.   

While Cen X-4 and Aql X-1 are clear examples of variable quiescent sources, other LMXBs are known to be extremely steady.  Consequently, they are likely reliable sources for measuring radii.  Among these are X7 in 47 Tuc \citep{heinke06} and the 4 quiescent neutron stars studied by \citet{guillot13} that have multiple observations. 

In this paper, we look at quiescent neutron star LMXBs in globular clusters using archival {\it Chandra} observations. This is because globular clusters contain multiple quiescent neutron star LMXBs and because their distance is known to much better accuracy than most binary systems elsewhere in the galaxy. This reduces the uncertainty in the neutron star radius measurement.  Specifically, this paper focuses on establishing if quiescent variability is present in a sample of 9 targets, 5 in the globular cluster NGC 6440 (CX 1, 2, 3, 5 and 7) and 4 in the globular cluster Terzan 5 (CX 9, 12, 14 and 15). Note that throughout the paper we refer to the sources in NGC~6440 using their ID number from \citet{pooley02}, while for Terzan we use the ID number from \citet{heinke06b}. We analyze 4 archival {\it Chandra} observations of NGC 6440 and 8 archival {\it Chandra} observations of Terzan 5 (all publicly available observations as of summer 2013).  Details of the observations are given in Table~\ref{tab:obs}.  With the exception of CX1 in NGC 6440, the spectra of these sources have not been previously presented or examined for variability. Thus, our study significantly increases the number of quiescent LMXBs where variability has been searched for,  thus helping understand whether variability is a common feature among quiescent neutron star LMXBs.    

\begin{table}
\begin{center}
\small
\caption{{\it Chandra} observations analyzed here}
\label{tab:obs}
\renewcommand{\arraystretch}{1.2}
\begin{tabular}{cccc}
\hline

\multicolumn{1}{c}{Obs.}
&\multicolumn{1}{c}{Obs.}
&\multicolumn{1}{c}{Start Date}
&\multicolumn{1}{c}{Exposure Time}\\
\multicolumn{1}{c}{No.}
&\multicolumn{1}{c}{ID}
&&\multicolumn{1}{c}{(ks)}\\
\hline
\multicolumn{4}{c}{NGC 6440}\\
\hline
1
&947
&2000 Jul 04
&23.28\\
2
&3799
&2003 Jun 27
&24.05\\
3
&10060
&2009 Jul 28
&49.11\\
4
&11802
&2009 Aug 10
&4.91\\
\hline
\multicolumn{4}{c}{Terzan 5}\\
\hline
1
&3798
&2003 Jul 13
&39.34\\
2
&10059
&2009 Jul 15
&36.26\\
3
&12454
&2011 Nov 03
&9.84\\
4
&13225
&2011 Feb 17
&29.67\\
5
&13252
&2011 Apr 29
&39.54\\
6
&13705
&2011 Nov 05
&13.87\\
7
&14339
&2011 Nov 08
&34.06\\
8
&13706
&2012 May 13
&46.46\\
\hline
\end{tabular}
\end{center}
Note --- all observations were performed using the ACIS-S instrument in FAINT mode with the exception of observation 4 of NGC 6440, which was made using the ACIS-S instrument in VFAINT mode.
\end{table}

\subsection{NGC 6440}

NGC 6440 is a globular cluster at a distance of $8.5\pm0.4$~kpc, with extinction of $E(B-V) = 1.0$ \citep{ortolani94}.  Previous {\it Chandra} studies of this cluster identified 24 X-ray sources with a 0.5 -- 2.5 keV luminosity above $\sim 2\times10^{31}$ erg s$^{-1}$ within the cluster's half-mass radius \citep{pooley02}.  Of those sources, 8 are quiescent neutron star candidates \cite{heinke03b}.  We selected the 5 brightest of those (CX1, 2, 3, 5, 7) as the others are too faint for reliable spectral fitting.  There are two known X-ray transients in NGC 6440.  The first, SAX~J1748.9$-$2021, was seen to go into outburst in 1998, 2001 and 2005 \citep{intzand99,intzand01,markwardt05}.  Timing analysis revealed SAX~J1748.9$-$2021 to be an intermittent accreting millisecond pulsar \citep{altamirano08}.  The 2001 outburst fell between the first two {\it Chandra} observations of NGC 6440, and the 2005 outburst occurred between {\it Chandra} observations 2 and 3.  \citet{pooley02} identified CX1 as the quiescent counterpart to SAX~J1748.9$-$2021.  Analysis of the first two {\it Chandra} observations of CX1 showed that it was variable in quiescence, with the variability likely due to the power-law component \citep{cackett05}.

In 2009 a second transient X-ray source in NGC 6440 was detected \citep{heinke10}, officially called CXOG1b~J174852.7$-$202124, but often referred to as NGC 6440 X-2 \citep[not to be confused with CX2 from][]{pooley02}.  This transient is too dim to be detected in quiescence, and consequently we do not include it in our analysis.  However, during outburst, millisecond pulsations were detected \citep{altamirano10}. The 2009 {\it Chandra} observation of NGC 6440 was taken during the outburst of CXOG1b~J174852.7$-$202124.  Despite this, all four quiescent sources were detected.

Of the 5 quiescent objects in NGC 6440 we study here, only CX1 is a confirmed neutron star that has been seen in outburst, while CX 2, 3, 5 and 7 are instead all candidate quiescent neutron stars.  

\subsection{Terzan 5}

The globular cluster Terzan 5 is at a distance of  5.5 $\pm$ 0.9 kpc \citep{ortolani07}.  Previous {\it Chandra} studies of this cluster identified 13 of the 50 X-ray sources detected down to a $1 - 6$ keV X-ray luminosity of $3.1\times10^{31}$ erg s$^{-1}$ as quiescent neutron star candidates \citep{heinke03a,heinke03b,heinke06b}.  There have been 3 X-ray transients detected in Terzan 5.  However, none of those sources are part of this current study.  The first X-ray transient in Terzan 5 is EXO~1745$-$248 \citep[CX3 in][]{heinke06b}.  It was found by \citet{wijnands05} to show a hard quiescent spectrum with no thermal component, thus we do not consider it here.  The second is IGR~J17480$-$2446 \citep[CX25 in][]{heinke06b}.  It is a crustal cooling source, and thus its thermal evolution has been studied in detail already \citep{degenaar11a,degenaar11b,degenaar11c,degenaar13}.  Furthermore, the quiescent variability of the third X-ray transient, Swift~J174805.3-244637 \citep[CX2 in][]{heinke06b} has been studied by \citet{bahramian14} who find that although it is variable during quiescence, the variability is attributable to the non-thermal component (and hence the thermal component is steady).

We study the 4 brightest remaining quiescent neutron star candidates (CX 9, 12, 14 and 15), none of which are known transients.  During the third {\it Chandra} observation of Terzan 5, CX9 and CX12 were not detected due to the outburst of CX3.  Aside from that, all 4 sources were detected in all {\it Chandra} observations.

\section{Data Reduction}

The data were reprocessed with the \emph{Chandra} software CIAO v. 4.5 with the calibration database CALDB v. 4.5.6 using the \texttt{chandra\_repro} script.  The \texttt{specextract} tool was used to extract appropriate spectral files from source and background regions. We use circular source extraction regions with a radius of 1.5\arcsec\ centered on the source positions in \citet{heinke03b} and \citet{heinke06b} for NGC 6440 and Terzan 5, respectively.  Background extraction regions for observations 1, 2, and 4 of NGC 6440 were annuli with an inner radius of 17\arcsec\ and outer radius of 28\arcsec.  For observation 3 of NGC 6440 (where the source CXOG1b J174852.7-202124 was in outburst), circular background extraction regions of 3\arcsec\ for CX1, and 11\arcsec\ for CX2, CX3, CX5, and CX7 were selected to include the background close to the source and avoid the outburst emission of CXOG1b J174852.7-202124.  For sources in Terzan 5, 20\arcsec\ circular background regions close to the source were selected for all observations since an annular region that did not contain contamination from other sources was not possible.

\
\section{Spectral Analysis}

\begin{figure}
\centering
\includegraphics[angle=270,width=8cm]{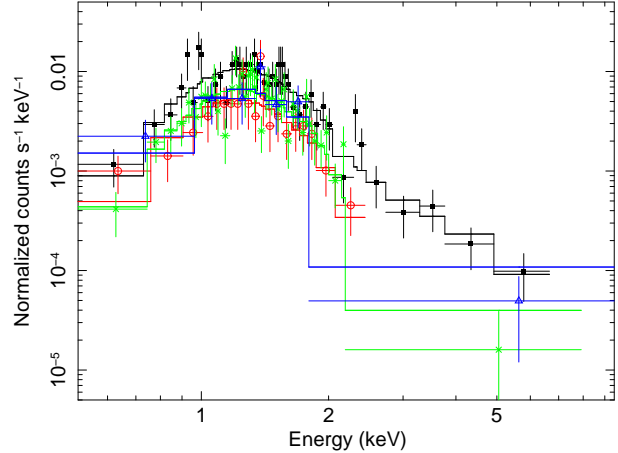}
\caption{The quiescent X-ray spectra of CX1 in NGC 6440.   Observation 1  is shown as black squares, observation 2 as red circles, observation 3 as green crosses, and observation 4 as blue triangles.  The solid lines through the data points are the fitted models for each observation.  The spectra are visually rebinned for clarity. }
\label{fig:spectracx1}
\end{figure}

Spectral fits were performed in the $0.5- 10$ keV energy range using XSPEC v. 12.8.0 \citep{arnaud96}, first using the C-statistic \citep{cash79}, detailed in Section~\ref{subsec:cstat}, and then $\chi^2$ statistics, detailed in Section~\ref{subsec:chisquared}.  The C-statistic was used to include even the faintest observations, since at least one observation of each source had too few counts to use $\chi^2$ statistics.   Fits using $\chi^2$ statistics were included to compare models with an easy to interpret goodness of fit measure, which the C-statistic does not provide.

In order to test the importance of the assumed distance on the spectral fit parameters, we performed spectral fits to the highest signal-to-noise ratio observations of the brightest objects in both NGC 6440 and Terzan 5.  We fitted the spectra with an absorbed neutron atmosphere model, \texttt{phabs} in order to account for galactic absorption in the direction of the source and \texttt{nsatmos} \citep{heinke06a}, assuming a mass of 1.4 $M_\odot$ and a radius of 10 km. The normalization of the hydrogen atmosphere model was fixed to 1, as the entire neutron star surface is assumed to be emitting.  We also included a power-law component, \texttt{pow}, to fit the non-thermal component.  

We fit observation 1 of CX1 in NGC 6440 assuming distances of 8.1, 8.5 and 8.9 kpc (exploring the full 1$\sigma$ uncertainty in the distance).  The best-fit parameters were all consistent within the uncertainties.  We fit observation 8 of CX 9 in Terzan 5 assuming distances of 4.6, 5.5 and 6.4 kpc.  Again, the best-fit parameters were all consistent within the uncertainties.  Given this, we proceeded by assuming a distance of 8.5 kpc for NGC 6440 and 5.5 kpc for Terzan 5 for all other spectral fits.

We continued by fitting all observations of each source simultaneously, tying the column density between observations.  As before, we assumed a 1.4 M$_{\odot}$, 10 km radius neutron star with the entire surface emitting. All uncertainties throughout this article are given at the 1$\sigma$ level.  The unabsorbed thermal flux and the unabsorbed total flux were calculated using the \texttt{cflux} model for the 0.5 -- 10 keV energy range. 

In order to assess the level of variability between observations we calculate the root mean square variability amplitude, $F_{var}$, following \citet{vaughan03} using the thermal and unabsorbed fluxes from \texttt{cflux}.  Tables~\ref{tab:ngc6440b} gives $F_{var}$ for the unabsorbed thermal and unabsorbed total flux for each source.  We quote upper limits for $F_{var}$ when the calculated number is consistent with zero within 1$\sigma$.  Where no number is quoted, the calculation of $F_{var}$ failed because of lack of detectable variability, as was the case for all sources in Terzan 5 when fit using C-statistics.

\subsection{Spectral Analysis Using C-statistics}
\label{subsec:cstat}

All observations were grouped to one count per bin, as is recommended for the C-statistic.  All observations of each source were fit simultaneously with the column density tied between observations.  We assumed a canonical mass and radius of 1.4 M$_{\odot}$ and 10 km, respectively, with the whole neutron star surface emitting.  For the power-law component,  only in the brightest source, CX1 in NGC 6440, were we able to constrain the power-law index.  For all other sources we therefore choose to fix the power-law index at 1.5, similar to the value seen in Cen X-4 \citep[e.g.][]{cackett10, cackett13_cenx4,bernardini13}.  The effective temperature was left as a free parameter to allow for careful examination of thermal variability.  

In at least one observation of each source in NGC 6440 the power-law normalization was consistent with zero within 2$\sigma$ and there was no significant detection above 3 keV.  However, it is not clear whether this is due to the lower signal-to-noise ratio in those observations or the true disappearance of the power-law component.  We therefore compared the temperatures with and without the power-law component for these cases, and found that all the temperatures were consistent within 1$\sigma$. When it is not required (when the power-law normalization is consistent with zero) we therefore give the spectral fit parameters without the power-law component, and simply quote the 1$\sigma$ upper limit on the power-law normalization. All spectral parameters for NGC 6440 are given in Table  \ref{tab:ngc6440b}, and as an example, we show the spectra of CX1 in Figure~\ref{fig:spectracx1}.   

We follow the same fitting procedure (described above for NGC 6440) for the Terzan 5 sources, however, in these cases we always fit with the power-law normalization as a free parameter.  For Terzan 5 we found that the power-law component significantly improved the fit for all observations of all sources.  Although we cannot use an F-test with the C-statistic, we find that the power-law normalization is always significantly greater than zero, indicating that the parameter is required. The spectral fit parameters for Terzan 5 are given in Table~\ref{tab:terz}.  

The column density was a free parameter for all observations of all sources in both clusters and was consistent within $2\sigma$ for sources in the same cluster.  Such deviations are expected given the large number of spectra examined.\\

\subsection{Spectral Analysis Using $\chi^2$ Statistics}
\label{subsec:chisquared}
 
All observations were grouped to 15 counts per bin as is required to use $\chi^{2}$ statistics.  Due to the low number of counts in observation 4 for all objects, only observations 1- 3 could be analyzed for CX1, CX2, and CX3 in NGC 6440.  Only observation 3 could be used for analysis for CX5 and CX7 in NGC 6440.  Since our aim was to investigate variability between observations, these two sources were not included in this analysis.  In Terzan 5, we were only able to test variability for Observations 1, 2, 4, 5, 7, and 8 of CX9 and CX12.   All other observations had too few counts to allow for analysis with $\chi^2$ statistics.

To compare models, we followed a three-step fitting procedure.  In all fits, the column density was tied between observations, and a mass of 1.4 M$_{\odot}$ and a neutron star radius of 10 km was assumed.  The entire surface was taken to be emitting.  First, we tied all parameters between observations.  The second fit allowed only the power-law normalization to vary.  The third fit allowed only the effective temperature to vary.  When an unphysical value was given for the power-law index, it was fixed to 1.5.

Spectral parameters for CX1 in NGC 6440 are given in Table~\ref{tab:cx1bah}.  Spectral fits for CX2 and CX3 are discussed in Section~\ref{sec:Results}.  Spectral fits for CX9 and CX12 in Terzan 5 are given in Table~\ref{tab:t5bah} and discussed in Section \ref{sec:Results}.

\begin{figure}
\begin {flushleft}
~~~~~~~~~~\textbf{NGC 6440}
\end{flushleft}
\centering
\includegraphics[width=8.0cm]{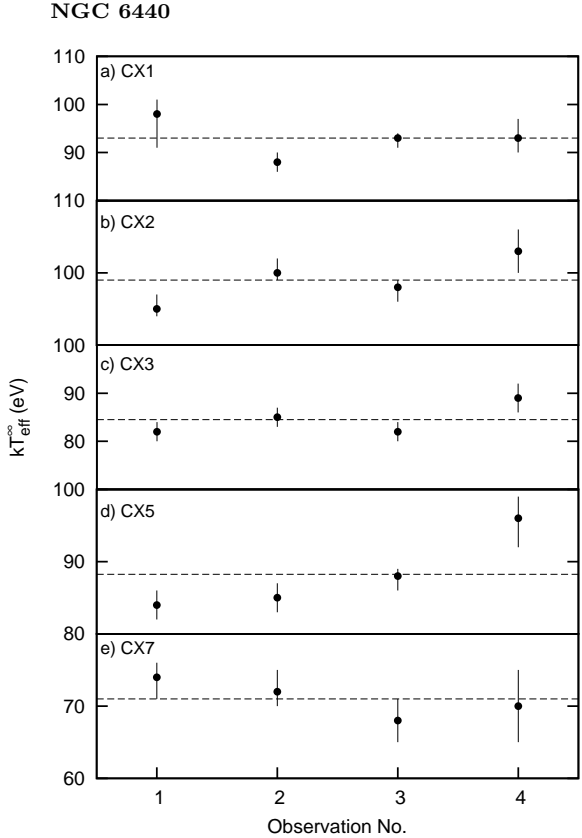}
\caption{The effective temperature and 1$\sigma$ error bars are plotted for each observation of each source in NGC 6440 for a) CX1, b) CX2, c) CX3, d) CX5, and e) CX7.  The dashed lines indicate the average effective temperature for each source.}
\label{fig:ngc_ktb}
\end{figure}

\begin{figure}
\begin {flushleft}
~~~~~~~~~~\textbf{NGC 6440}
\end{flushleft}
\centering
\includegraphics[width=8.0cm]{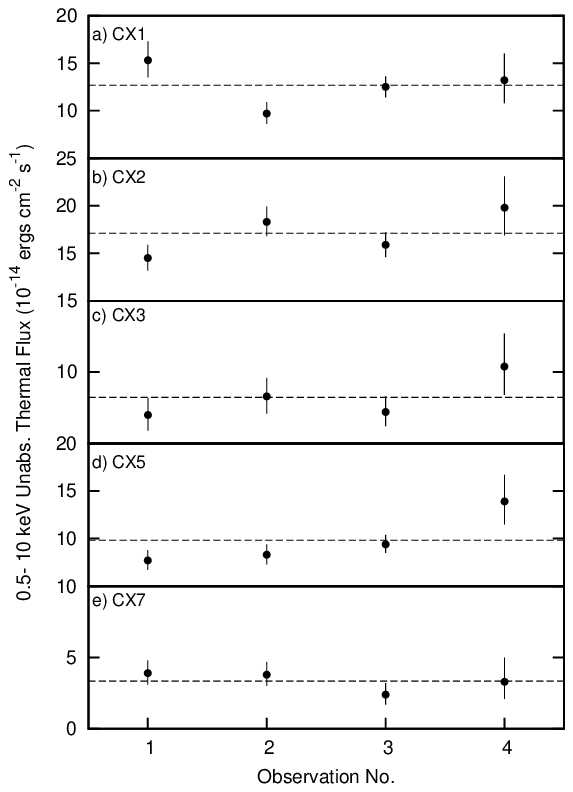}
\caption{The 0.5 -- 10 keV unabsorbed thermal flux for each observation of each source in NGC 6440 for a) CX1, b) CX2, c) CX3, d) CX5, and e) CX7.  The dashed lines indicate the average thermal flux for each source.}
\label{fig:ngc_thflb}
\end{figure}

\begin{figure}
\begin {flushleft}
~~~~~~~~~~\textbf{NGC 6440}
\end{flushleft}
\centering
\includegraphics[width=8.0cm]{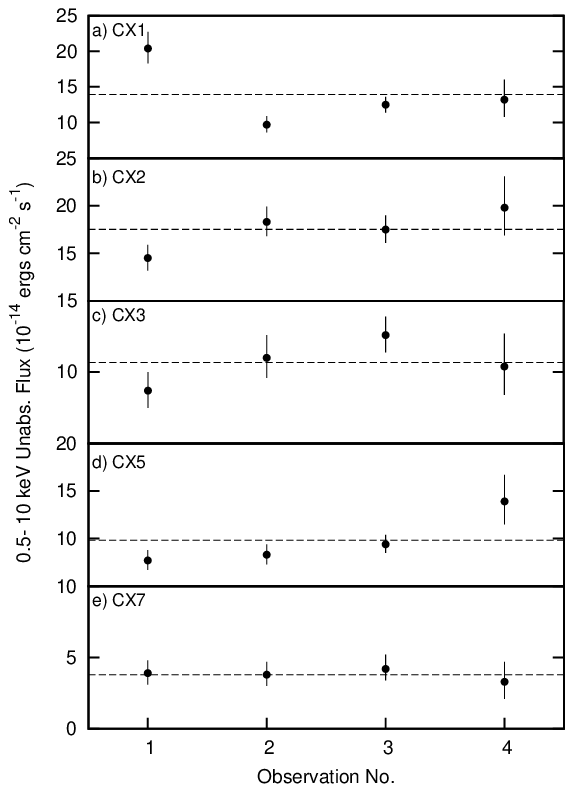}
\caption{The 0.5 -- 10 keV unabsorbed flux for each observation of each source in NGC 6440 for a) CX1, b) CX2, c) CX3, d) CX5, and e) CX7.  The dashed lines indicate the average flux for each source.}
\label{fig:ngc_flb}
\end{figure}

\section{Results}
\label{sec:Results}

\begin{figure*}
\begin{center}
\begin{tabular}{cc}
\includegraphics[angle=270,width=0.35\linewidth]{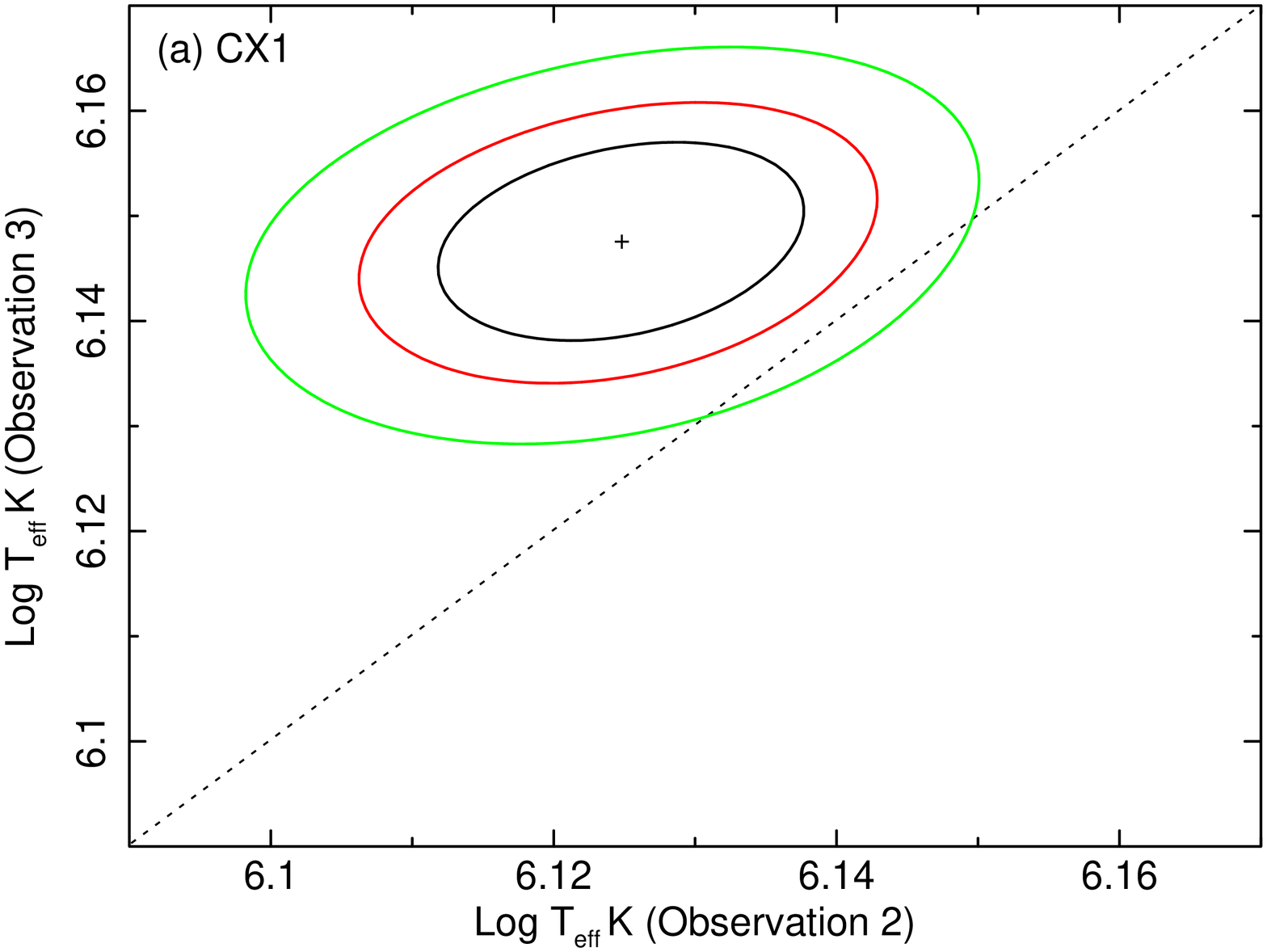}
&\includegraphics[angle=270,width=0.35\linewidth]{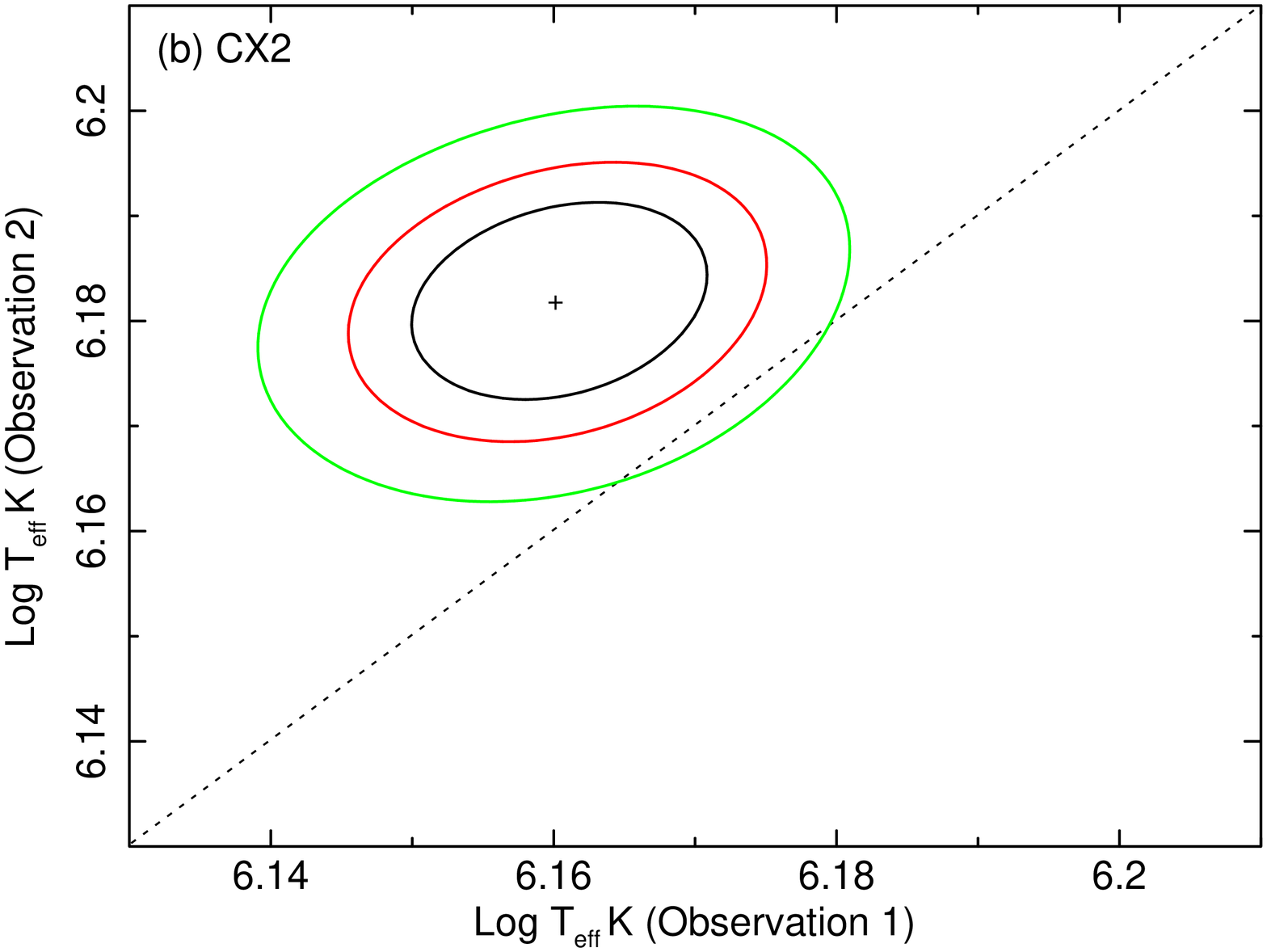}\\
\includegraphics[angle=270,width=0.35\linewidth]{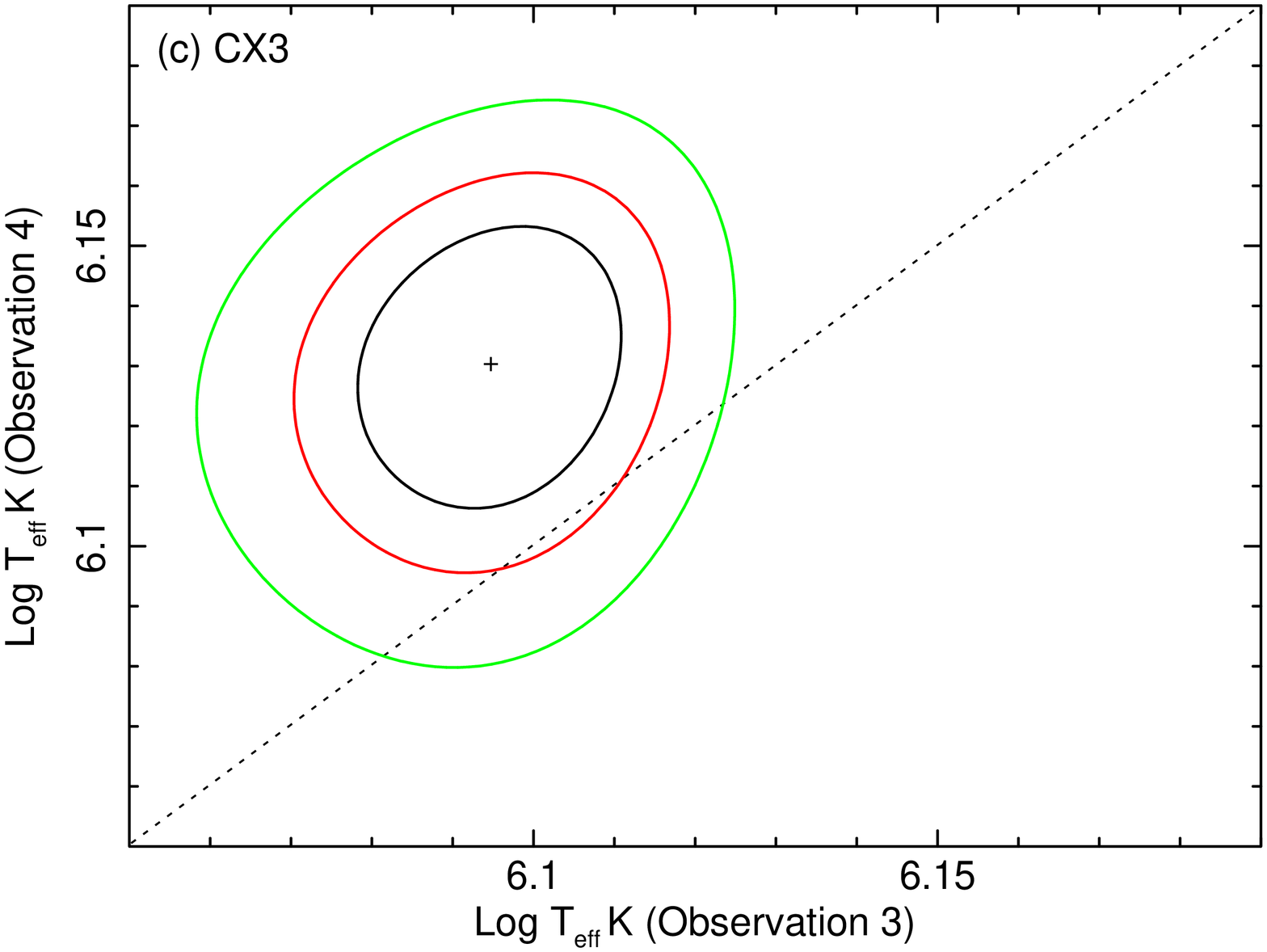}
&\includegraphics[angle=270,width=0.35\textwidth]{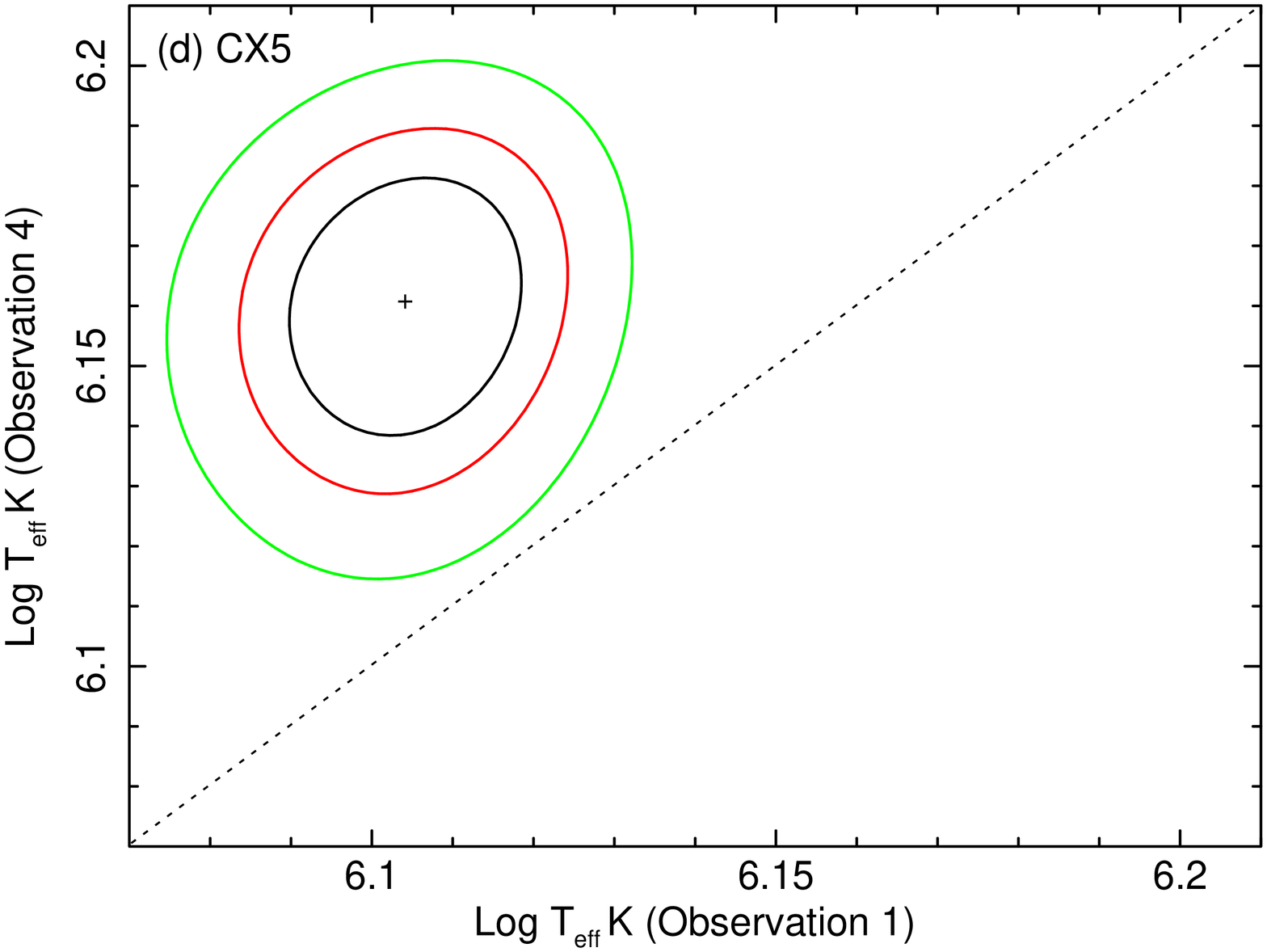}\\
\includegraphics[angle=270,width=0.35\textwidth]{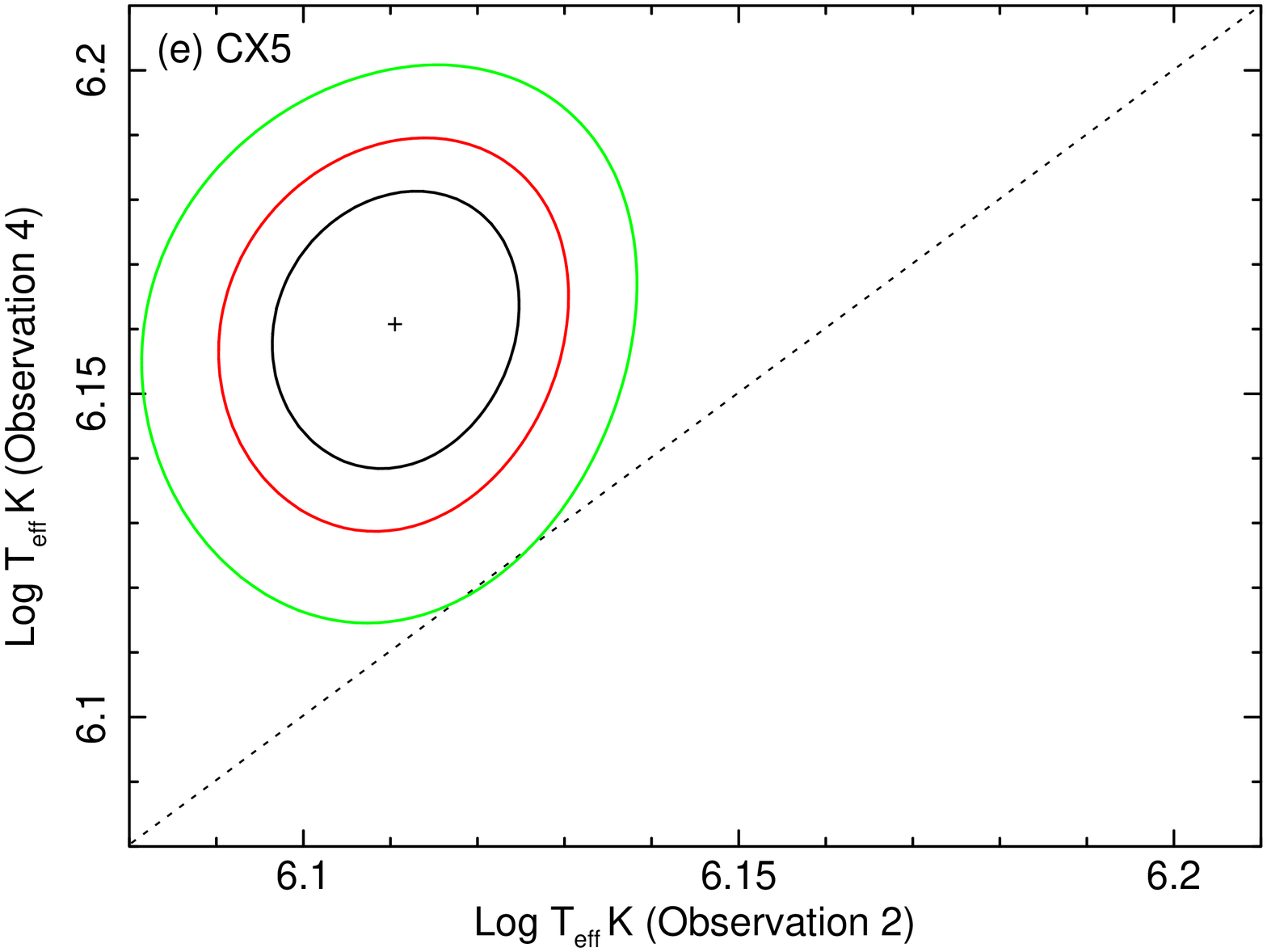}
&\includegraphics[angle=270,width=0.35\textwidth]{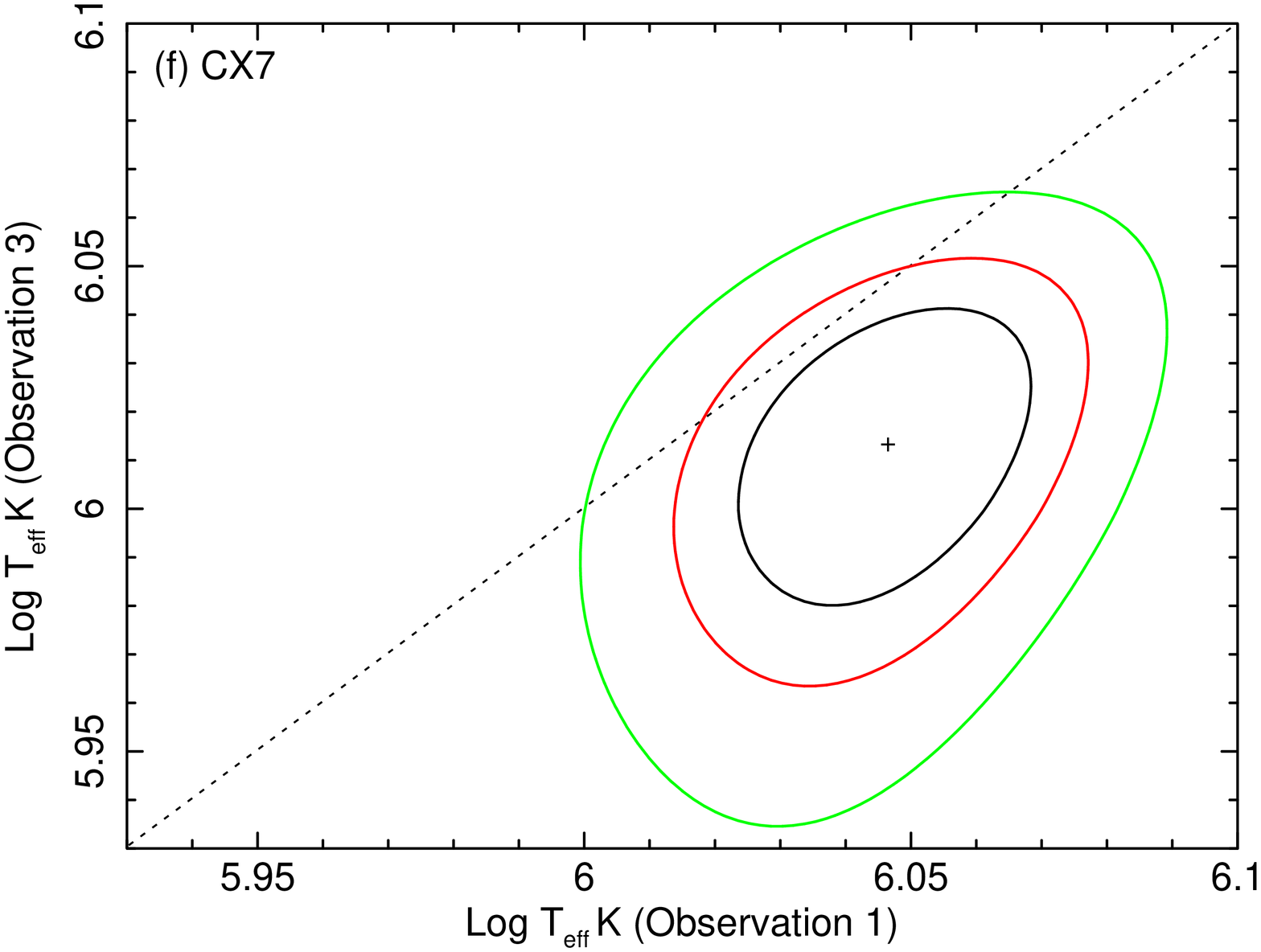}\\
\end{tabular}
\end{center}
\caption{Comparison of $Log~T_{\mathit{eff}}$ K for observations between which there is the most variability in temperature for each source in NGC 6440 when fit using C-statistics.  The innermost contour lines represent the boundary of the 68\% confidence level, the middle the 90\% confidence level, and the outermost the 99\% confidence level with a dashed line of equal temperature.  (a) Observations 2 and 3 of CX1.  (b) Observations 1 and 2 of CX2.  (c)  Observations 3 and 4 of CX3.  (d)  Observations 1 and 4 of CX5.  (e) Observations 2 and 4 of CX7.  $Log~T_{\mathit{eff}}$ differs between observations 1 and 4 of CX5 at greater than the 99\% confidence level.  The next most variable comparison of effective temperature between observations of CX5, observations 2 and 4, are consistent at better than the 99\% confidence level.  All other comparisons of $Log~T_{\mathit{eff}}$ shown are consistent within the 99\% confidence level or better.}
\label{fig:ngc_compktb}
\end{figure*}

\begin{figure}
\begin {flushleft}
~~~~~~~~~~\textbf{Terzan 5}
\end{flushleft}
\centering
\includegraphics[width=8.0cm]{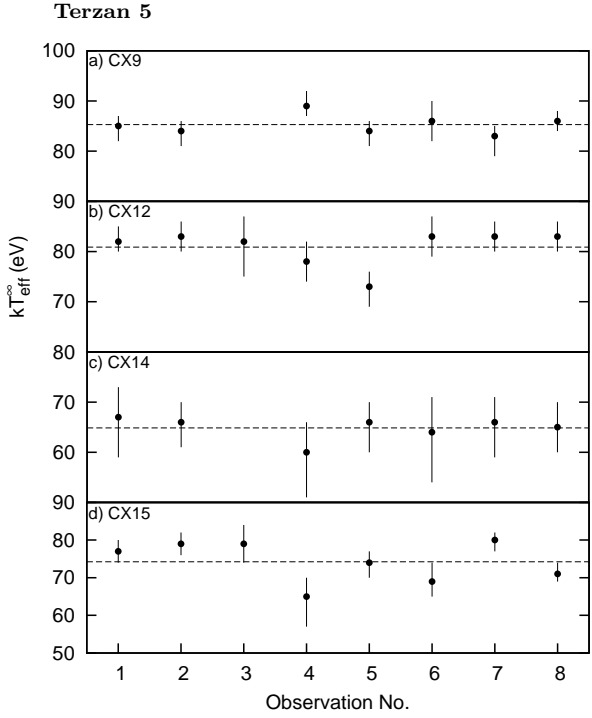}
\caption{The effective temperature and 1$\sigma$ error bars are plotted for each observation of each source in Terzan 5 for a) CX9, b) CX12, c) CX14, and d) CX15.  The dashed lines indicate the average effective temperature for each source.}
\label{fig:terz_kt}
\end{figure}

\begin{figure}
\begin {flushleft}
~~~~~~~~~~\textbf{Terzan 5}
\end{flushleft}
\centering
\includegraphics[width=8.0cm]{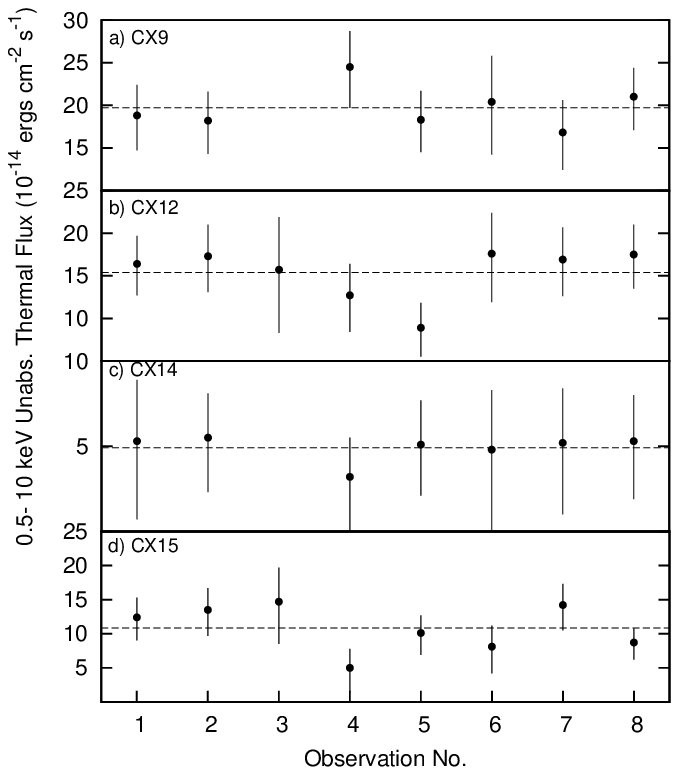}
\caption{The 0.5 -- 10 keV unabsorbed thermal flux for each observation of each source in Terzan 5 for a) CX9, b) CX12, c) CX14, and d) CX15.  The dashed lines indicate the average flux for each source.}
\label{fig:terz_thfl}
\end{figure}

\begin{figure}
\begin {flushleft}
~~~~~~~~~~\textbf{Terzan 5}
\end{flushleft}
\centering
\includegraphics[width=8.0cm]{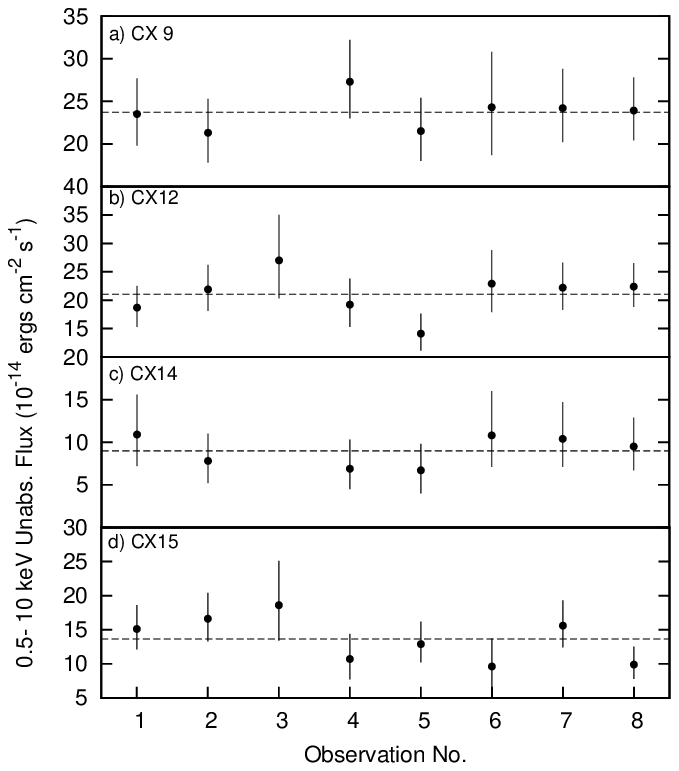}
\caption{The 0.5 -- 10 keV unabsorbed total flux for each observation of each source in Terzan 5 for a) CX9, b) CX12, c) CX14, and d) CX15.  The dashed lines indicate the average flux for each source.}
\label{fig:terz_fl}
\end{figure}

\begin{figure*}
\begin{center}
\begin{tabular}{cc}
\includegraphics[angle=270,width=0.35\linewidth]{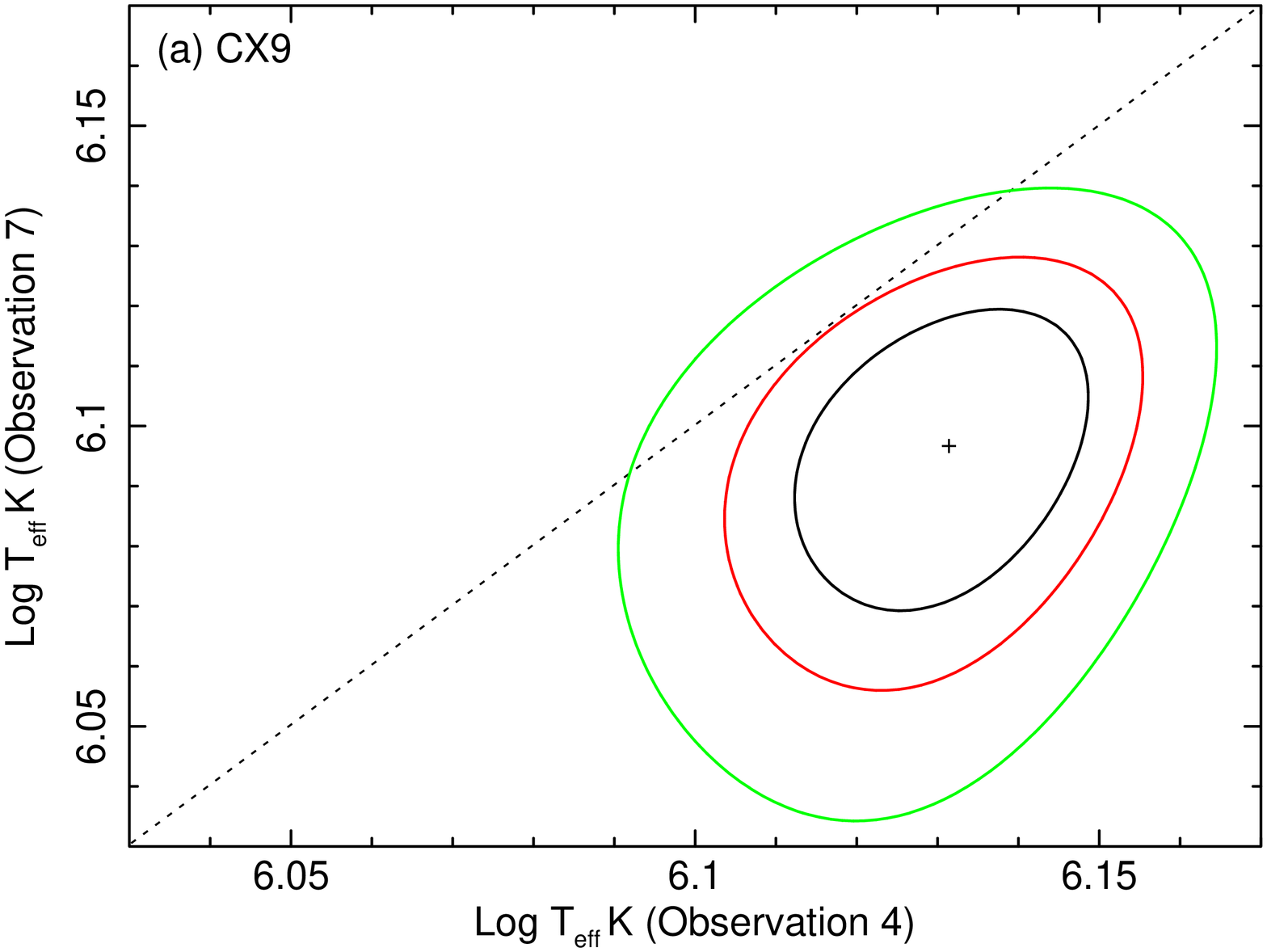}
&\includegraphics[angle=270,width=0.35\linewidth]{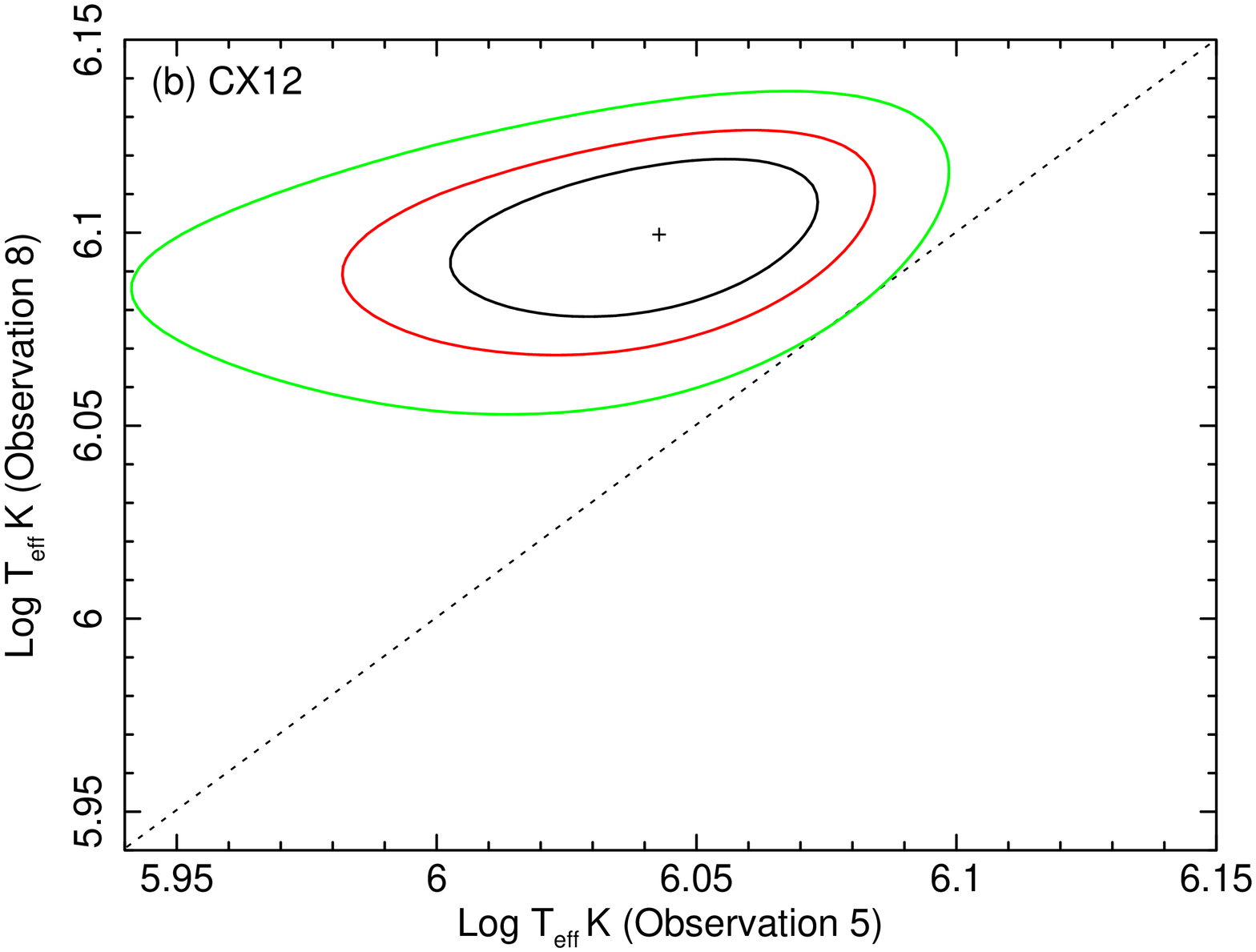}\\
\includegraphics[angle=270,width=0.35\linewidth]{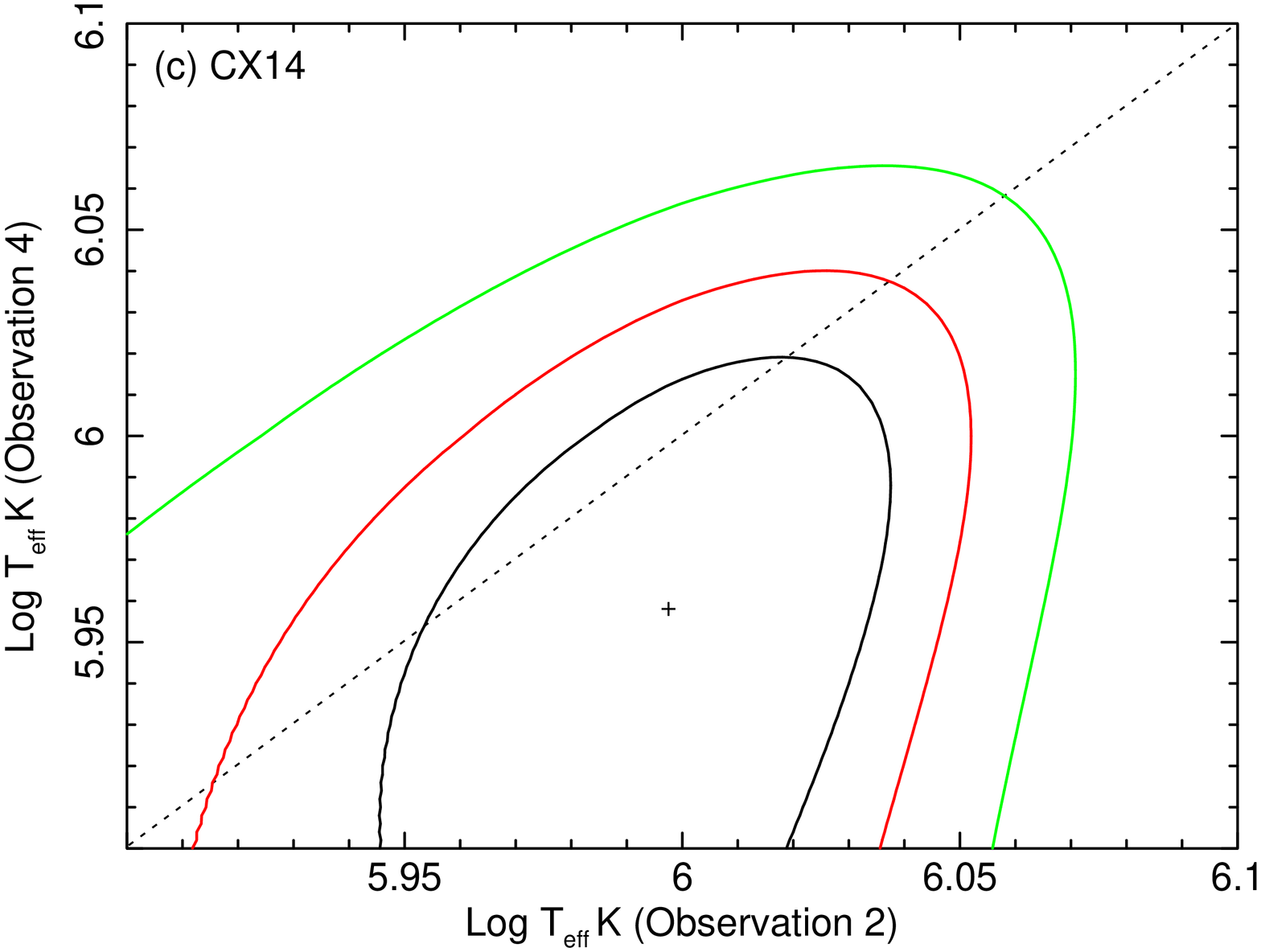}
&\includegraphics[angle=270,width=0.35\linewidth]{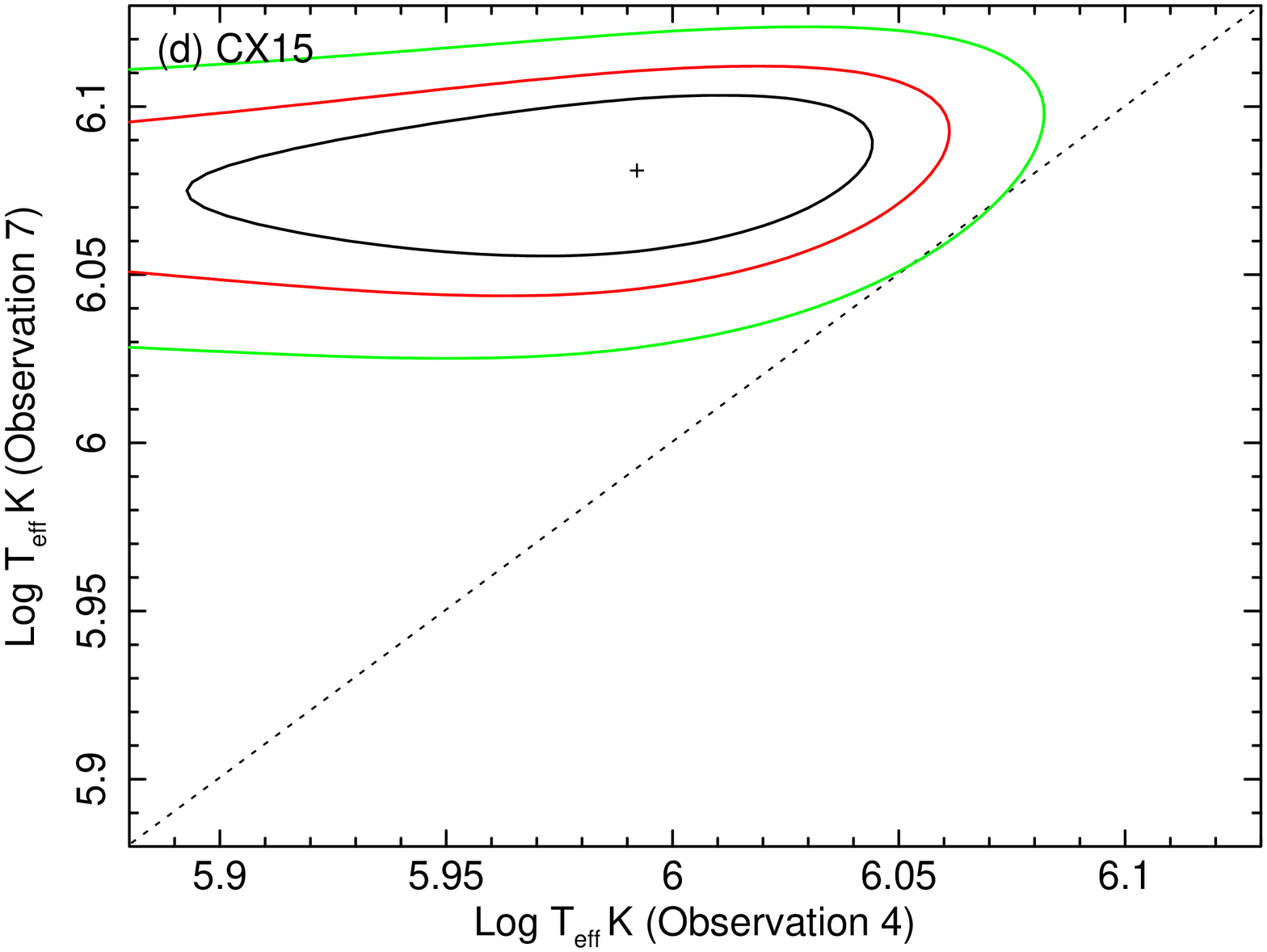}\\
\end{tabular}
\end{center}
\caption{Comparison of $Log~T_{\mathit{eff}}$ K for observations between which there is the most variability in temperature for each source in Terzan 5 when fit using C-statistics. The innermost contour lines represent the boundary of the 68\% confidence level, the middle the 90\% confidence level, and the outermost the 99\% confidence level with a dashed line of equal temperature.  (a) Observations 4 and 7 of CX9.  (b) Observations 5 and 8 of CX12.  (c) Observations 2 and 4 of CX14.  (d) Observations 4 and 7 of CX15.    All comparisons of $Log~T_{\mathit{eff}}$ shown are consistent at the 99\% confidence level or better.}
\label{fig:terz_compkt}
\end{figure*}

A plot of effective temperature for each source for each observation for the spectral fits using C-statistics is presented in Figure~\ref{fig:ngc_ktb} for NGC 6440, along with the 0.5 -- 10 keV unabsorbed thermal flux in Figure~\ref{fig:ngc_thflb} and the unabsorbed total flux for the same energy range in Figure~\ref{fig:ngc_flb}.   Figure~\ref{fig:ngc_compktb} shows the 68\% (1$\sigma$), 90\% (1.7$\sigma$), and 99\% (2.6$\sigma$) confidence level contours for the temperatures for the observations of NGC 6440 fit using C-statistics which show the most variation between them.  We show the same set of figures for Terzan 5 in Figures~\ref{fig:terz_kt}, \ref{fig:terz_thfl},  \ref{fig:terz_fl} and \ref{fig:terz_compkt}.

Observations 1 and 2 of CX1 in NGC 6440 were previously studied by \citet{cackett05}.  We agree with their findings that the power-law component is only statistically required for the first observation.  The two additional observations we study here (observations 3 and 4) also did not require a power-law component.    The effective temperature of CX1 was constant within the 99\% confidence level when fit using C-statistics. 

When the power-law normalization was allowed to vary freely, it was consistent with zero for every observation of CX5  when fit using C-statistics, and so the spectra were refit with only an absorbed neutron atmosphere model.  When fit without a power-law component, the effective temperature of CX5 was not found to be consistent between observations 1 and 4.  They differ at more than the 99\% confidence level.  On the other hand, for the next most variable pair of observations, the comparison between observations 2 and 4, the effective temperature was constant within the 99\% confidence level.  Observation 4 of NGC 6440 was 4.91 ks in duration, $< 1/4$ of the duration of the other observations of NGC 6440, resulting in few counts.  The low signal-to-noise ratio does not allow for determination of whether or not the spectrum is better fit with or without a power-law component.  

In all other cases in NGC 6640  when fitting with C-statistics, the effective temperature is always constant within the 99\% confidence level. The effective temperatures of CX3 and CX7 are found to be consistent within the 90\% confidence level as shown in Figure~\ref{fig:ngc_compktb}.  

When fit using $\chi^2$ statistics, only observations $1-3$ of CX1 in NGC 6440 could be included in spectral fitting since $\chi^2$ statistics require grouping data to 15 counts per bin and the signal to noise ratio was low for observation 4.  When fit with all parameters tied, the best fit value of the power-law normalization is zero; however, this gives a poor fit with a $\chi^2$ value of 77.65 for 36 degrees of freedom.  The spectral results for the fits with only the power-law normalization free to vary and for the fits with only the effective temp free to vary are reported in Table~\ref{tab:cx1bah}.   Allowing only the power-law normalization to vary gives a high value for this parameter of $4.08\times10^{-5}$ for the first observation, but only upper limits for observations 2 and 3.  This gives a good fit with a $\chi^2$ value of 34.42 for 34 degrees of freedom.  Allowing only the effective temperature to vary between observations, yields a power-law index with an unphysical best fit value of 7.6, and so it was fixed at 1.5.  The power-law normalization was $\le 0.7$, consistent with 0 and hence there is no statistical requirement for a power-law component when this source is fit with the effective temperature tied between observations.  The spectra were refit with only an absorbed neutron star atmosphere model.  The spectral fit parameters are given without the power-law component and we quote the $1\sigma$ upper limit of the power-law normalization.  The effective temperature was found to differ at greater than the 99\% confidence level between observations 1 and 2 and between observations 1 and 3, but was constant within the 90\% confidence level between observations 2 and 3.  This variability in the thermal component is not surprising, since the source count rate is variable, and the variability must be attributed to the only free parameter.  If we do not include a variable power-law component, the thermal component must vary.  This fit gives a $\chi^2$ value of 37.71 for 37 degrees of freedom.  The fit with the power-law normalization free gives a is marginally lower $\chi^2$ value.  An F-test gives a probability of 0.08 of obtaining this improvement by chance, so it is not strong evidence of a better fit. In conclusion, CX1 in NGC 6440 is equally well fit by a variable thermal component or a variable power-law.  Therefore we cannot exclude thermal variability in this source.

Only observations 1, 2, and 3 of CX2 and CX3 in NGC 6440 could be analyzed using $\chi^2$ statistics.  Fits with $\chi^2$ values of 45.77 for 43 degrees of freedom and 21.72 for 32 degrees of freedom, respectively, are given for these two sources when all parameters are tied.  Thus, there is no evidence for variability.

For the sources in Terzan 5, the effective temperature was constant within the 99\% confidence level for all sources using both fit statistics.  In particular, CX14 in Terzan 5 was constant within the 68\% confidence level as shown in Figure~\ref{fig:terz_compkt}, as were CX9 and CX12, when fit using $\chi^2$ statistics. 

CX9 in Terzan 5 was well fit when using $\chi^2$ statistics when only the power-law normalization was allowed to vary, with a $\chi^2$ value of 32.40 for 30 degrees of freedom.  Tying all parameters gave a poor fit with a $\chi^2$ value of 46.17 for 35 degrees of freedom, and allowing only the effective temperature to vary between observations gave a poor fit with a $\chi^2$ value of 42.95 for 30 degrees of freedom.  Tying all parameters of CX12 gave a poor fit with a $\chi^2$ statistic of 48.70 for 34 degrees of freedom.  Allowing only the power-law normalization to vary between observations gave a better fit with a $\chi^2$ value of 36.09 for 29 degrees of freedom.  The effective temperature was given as an upper limit, and the power-law index was $3.1\pm0.5$, an unusually soft value.  When using $\chi^2$ statistics, CX12 in Terzan 5 was best fit when the effective temperature was the only free parameter, with a $\chi^2$ value of 30.46 for 30 degrees of freedom; however, the effective temperature was constant within the 68\% confidence level between observations.  These fit are given in Table~\ref{tab:t5bah}.  

To measure  how tightly constrained the thermal variability is, we calculated the largest allowed deviation from the mean temperature for each object  when fit using C-statistics, by calculating the difference between the most extreme 1$\sigma$ error bar (i.e. the limit which is furthest from the mean) and the mean.  For the objects in NGC 6440 the average value for this was 4 eV or 5\%.  For Terzan 5 the temperatures are less well constrained, with maximum 1$\sigma$ deviations of 7 eV or 10\% on average.

The unabsorbed flux of CX1 is seen to be variable and has the highest excess variance of the sources we study here with C-statistics, $F_{var} = 0.30\pm0.07$.  The excess variance of the unabsorbed thermal flux is $F_{var} = 0.12\pm0.09$,  a small fraction of the the overall excess variance.  This demonstrates that the variability seen in CX1 can mostly be attributed to the power-law alone.

The unabsorbed flux of CX5 is seen to be variable when fit using C-statistics.  When fit with a power-law component with a freely varying normalization, however, $F_{var}$ is consistent with zero.  Fitting CX5 without a power-law component decreases uncertainty in the calculated unabsorbed thermal flux, resulting in an increase in the excess variance.  If the unabsorbed flux of CX5 is indeed variable, it varies thermally, but the low number of source counts in observation 4 of this source does not allow for a strong conclusion.  

The unabsorbed flux of all sources in Terzan 5 did not vary beyond what is expected from measurement error when fit using C-statistics.

\begin{table*}
\small
\caption{Spectral parameters for sources in NGC 6440 using C-statisitcs}
\label{tab:ngc6440b}
\begin{center}
\renewcommand{\arraystretch}{1.2}
\renewcommand{\tabcolsep}{0.25cm}
\begin{tabular}{lcccc}
\hline
\hline
\multicolumn{1}{l}{Model Parameter}
&\multicolumn{1}{c}{Obs. 1}
&\multicolumn{1}{c}{Obs. 2}
&\multicolumn{1}{c}{Obs. 3}
&\multicolumn{1}{c}{Obs. 4}\\
\hline
\multicolumn{5}{c}{CX1}\\
\hline
\multicolumn{1}{l}{$N_H(10^{22} $cm$^{-2})$}
&\multicolumn{4}{c}{$0.73 \pm  0.03$}\\
$kT_\mathit{eff}^{\infty}$ (eV)
&$98_{-7}^{+3} $
&$88\pm2$
&$93_{-2}^{+1}$
&$93_{-3}^{+4}$\\
Power-law index, $\Gamma$
&\multicolumn{4}{c}{$1.6\pm1.0$}\\
Power-law norm. $(10^{-6})$
&$6.7_{-5.3}^{+23.9}$
&$\le 8.4$
&$\le 2.3$
&$\le 3.4$\\
Unabs. Thermal Flux (0.5- 10 keV) $(10^{-14}$ ergs cm$^{-2}$ s$^{-1}$)
&$15.5_{-1.8}^{+2.0}$
&$9.7_{-1.1}^{+1.2}$
&$12.5_{-1.1}^{+1.2}$
&$13.2_{-2.4}^{+2.8}$\\
Thermal $\mathrm{F}_\mathrm{var}$
&\multicolumn{4}{c}{$0.12 \pm 0.09$}\\
Unabs. Flux (0.5- 10 keV) $(10^{-14}$ ergs cm$^{-2}$ s$^{-1}$)
&$20.4_{-2.1}^{+2.3}$
&$9.7_{-1.1}^{+1.2}$
&$12.5_{-1.1}^{+1.2}$
&$13.2_{-2.4}^{+2.8}$\\
$\mathrm{F}_\mathrm{var}$
&\multicolumn{4}{c}{$0.30 \pm 0.07$}\\
\hline
\multicolumn{5}{c}{CX2}\\
\hline
\multicolumn{1}{l}{$N_H(10^{22} $cm$^{-2})$}
&\multicolumn{4}{c}{$0.74\pm 0.03$}\\
$kT_\mathit{eff}^{\infty}$ (eV)
&$95_{-1}^{+2}$
&$100_{-1}^{+2}$
&$98_{-2}^{+1}$
&$103\pm 3$\\
Power-law index, $\Gamma$
&\multicolumn{4}{c}{$1.5$}\\
Power-law norm. $(10^{-6})$
&$\le 0.5$
&$\le 1.3$
&$2.1\pm 0.7$
&$\le 12.2$\\
Unabs. Thermal Flux (0.5- 10 keV) $(10^{-14}$ ergs cm$^{-2}$ s$^{-1}$)
&$14.5_{-1.3}^{+1.4}$
&$18.3_{-1.5}^{+1.6}$
&$15.9\pm1.3$
&$19.8_{-2.9}^{+3.3}$\\
Thermal $\mathrm{F}_\mathrm{var}$
&\multicolumn{4}{c}{$\le 0.16$}\\
Unabs. Flux (0.5- 10 keV) $(10^{-14}$ ergs cm$^{-2}$ s$^{-1}$)
&$14.5_{-1.3}^{+1.4}$
&$18.3_{-1.5}^{+1.6}$
&$17.5_{-1.4}^{+1.5}$
&$19.8_{-2.9}^{+3.3}$\\
$\mathrm{F}_\mathrm{var}$
&\multicolumn{4}{c}{$\le 0.16$}\\
\hline
\multicolumn{5}{c}{CX3}\\
\hline
\multicolumn{1}{l}{$N_H(10^{22} $cm$^{-2})$}
&\multicolumn{4}{c}{$0.59\pm 0.04$}\\
$kT_\mathit{eff}^{\infty}$ (eV)
&$82\pm 2$
&$85\pm 2$
&$82\pm 2$
&$89\pm 3$\\
Power-law index, $\Gamma$
&\multicolumn{4}{c}{$1.5$}\\
Power-law norm. $(10^{-6})$
&$2.1\pm 0.8$
&$3.5\pm 1.0$
&$6.9\pm 0.9$
&$\le 1.3$\\
Unabs. Thermal Flux (0.5- 10 keV) $(10^{-14}$ ergs cm$^{-2}$ s$^{-1}$)
&$7.0_{-1.1}^{+1.2}$
&$8.3_{-1.2}^{+1.3}$
&$7.2_{-1.0}^{+1.1}$
&$10.4_{-2.0}^{+2.3}$\\
Thermal $\mathrm{F}_\mathrm{var}$
&\multicolumn{4}{c}{$\le 0.26$}\\
Unabs. Flux (0.5- 10 keV) $(10^{-14}$ ergs cm$^{-2}$ s$^{-1}$)
&$8.7_{-1.2}^{+1.3}$
&$11.0_{-1.4}^{+1.6}$
&$12.6_{-1.2}^{+1.3}$
&$10.4_{-2.0}^{+2.3}$\\
$\mathrm{F}_\mathrm{var}$
&\multicolumn{4}{c}{$\le 0.31$}\\
\hline
\multicolumn{5}{c}{CX5}\\
\hline
\multicolumn{1}{l}{$N_H(10^{22} $cm$^{-2})$}
&\multicolumn{4}{c}{$0.70\pm 0.04$}\\
$kT_\mathit{eff}^{\infty}$ (eV)
&$84\pm 2$
&$85\pm 2$
&$88_{-2}^{+1}$
&$96_{-4}^{+3}$\\
Power-law index, $\Gamma$
&\multicolumn{4}{c}{$1.5$}\\
Power-law norm. $(10^{-6})$
&$\le 1.1$
&$\le 1.9$
&$\le 1.1$
&$\le 9.1$\\
Unabs. Thermal Flux (0.5- 10 keV) $(10^{-14}$ ergs cm$^{-2}$ s$^{-1}$)
&$7.7_{-1.0}^{+1.1}$
&$8.3_{-1.0}^{+1.1}$
&$9.4_{-0.9}^{+1.0}$
&$13.9_{-2.4}^{+2.8}$\\
Thermal $\mathrm{F}_\mathrm{var}$
&\multicolumn{4}{c}{$0.24 \pm 0.09$}\\
Unabs. Flux (0.5- 10 keV) $(10^{-14}$ ergs cm$^{-2}$ s$^{-1}$)
&$7.7_{-1.0}^{+1.1}$
&$8.3_{-1.0}^{+1.1}$
&$9.4_{-0.9}^{+1.0}$
&$13.9_{-2.4}^{+2.8}$\\
$\mathrm{F}_\mathrm{var}$
&\multicolumn{4}{c}{$0.24 \pm 0.09$}\\
\hline
\multicolumn{5}{c}{CX7}\\
\hline
\multicolumn{1}{l}{$N_H(10^{22} $cm$^{-2})$}
&\multicolumn{4}{c}{$0.69\pm 0.07$}\\
$kT_\mathit{eff}^{\infty}$ (eV)
&$74_{-3}^{+2}$
&$72_{-2}^{+3}$
&$68\pm 3$
&$70\pm 5$\\
Power-law index, $\Gamma$
&\multicolumn{4}{c}{$1.5$}\\ 
Power-law norm. $(10^{-6})$
&$\le 1.0$
&$\le 2.5$
&$2.3\pm 0.6$
&$\le 2.7$\\
Unabs. Thermal Flux (0.5- 10 keV) $(10^{-14}$ ergs cm$^{-2}$ s$^{-1}$)
&$3.9_{-0.8}^{+0.9}$
&$3.8_{-0.8}^{+0.9}$
&$2.4_{-0.7}^{+0.8}$
&$3.3_{-1.2}^{+1.7}$\\
Thermal $\mathrm{F}_\mathrm{var}$
&\multicolumn{4}{c}{---}\\
Unabs. Flux (0.5- 10 keV) $(10^{-14}$ ergs cm$^{-2}$ s$^{-1}$)
&$3.9_{-0.8}^{+0.9}$
&$3.8_{-0.8}^{+0.9}$
&$4.2_{-0.8}^{+1.0}$
&$3.3_{-1.2}^{+1.7}$\\
$\mathrm{F}_\mathrm{var}$
&\multicolumn{4}{c}{---}\\
\hline

\end{tabular}
\end{center}
NOTE.---A mass of 1.4 M$_{\odot}$ and radius of 10 km was assumed for the \texttt{nsatmos} model.  
\end{table*}
\begin{table*}
\begin{center}
\small

\caption{Spectral parameters for sources in Terzan 5 using C-statistics}
\label{tab:terz}
\renewcommand{\arraystretch}{1.2}
\renewcommand{\tabcolsep}{0.15cm}
\begin{tabular}{lcccccccc}
\hline
\hline
\multicolumn{1}{l}{Model Parameter}
&\multicolumn{1}{c}{Obs. 1}
&\multicolumn{1}{c}{Obs. 2}
&\multicolumn{1}{c}{Obs. 3}
&\multicolumn{1}{c}{Obs. 4}
&\multicolumn{1}{c}{Obs. 5}
&\multicolumn{1}{c}{Obs. 6}
&\multicolumn{1}{c}{Obs. 7}
&\multicolumn{1}{c}{Obs. 8}\\
\hline
\multicolumn{9}{c}{CX9}\\
\hline
\multicolumn{1}{l}{$N_H$}
&\multicolumn{8}{c}{$2.10\pm 0.10$}\\
$kT_\mathit{eff}^{\infty}$
&$85_{-3}^{+2}$
&$84_{-3}^{+2}$
&---
&$89_{-2}^{+3}$
&$84_{-3}^{+2}$
&$86\pm 4$
&$83_{-4}^{+2}$
&$86\pm 2$\\
\multicolumn{1}{l}{$\Gamma$}
&\multicolumn{8}{c}{$1.5$}\\
P-l norm.
&$5.9\pm 1.1$
&$3.9\pm 1.0$
&---
&$3.5\pm 1.1$
&$4.2\pm 1.0$
&$4.8\pm 2.0$
&$9.3\pm 1.6$
&$3.7\pm 0.8$\\
Unabs. Thermal Flux
&$18.8_{-3.6}^{+4.1}$
&$18.2_{-3.4}^{+3.9}$
&---
&$24.5_{-4.2}^{+4.8}$
&$18.3_{-3.4}^{+3.8}$
&$20.4_{-5.4}^{+6.2}$
&$16.8_{-3.8}^{+4.4}$
&$21.0_{-3.4}^{+3.9}$\\
Unabs. Flux
&$23.5_{-3.7}^{+4.2}$
&$21.3_{-3.5}^{+4.0}$
&---
&$27.3_{-4.3}^{+4.9}$
&$21.5_{-3.5}^{+3.9}$
&$24.3_{-5.6}^{+6.5}$
&$24.2_{-4.0}^{+4.6}$
&$23.9_{-3.5}^{+3.9}$\\
\hline
\multicolumn{9}{c}{CX12}\\
\hline
\multicolumn{1}{l}{$N_H$}
&\multicolumn{8}{c}{$1.96\pm 0.12$}\\
$kT_\mathit{eff}^{\infty}$
&$82_{-2}^{+3}$
&$83\pm 3$
&$82_{-7}^{+5}$
&$78\pm 4$
&$73_{-4}^{+3}$
&$83\pm 4$
&$83\pm 3$
&$83\pm 3$\\
\multicolumn{1}{l}{$\Gamma$}
&\multicolumn{8}{c}{$1.5$}\\
P-l norm.
&$2.9\pm 0.8$
&$5.9\pm 1.2$
&$14.2\pm 3.2$
&$8.2\pm 1.9$
&$6.5\pm 1.0$
&$6.8\pm 2.0$
&$6.7\pm 1.2$
&$6.2\pm 1.1$\\
Unabs. Thermal Flux
&$16.4_{-3.3}^{+3.7}$
&$17.3_{-3.7}^{+4.2}$
&$15.7_{-6.2}^{+7.4}$
&$12.7_{-3.7}^{+4.3}$
&$8.9_{-2.9}^{+3.4}$
&$17.6_{-4.8}^{+5.7}$
&$16.9_{-3.8}^{+4.3}$
&$17.5_{-3.5}^{+4.0}$\\
Unabs. Flux
&$18.7_{-3.4}^{+3.8}$
&$21.9_{-3.8}^{+4.3}$
&$27.0_{-6.7}^{+8.0}$
&$19.2_{-3.9}^{+4.6}$
&$14.1_{-3.0}^{+3.5}$
&$22.9_{-5.0}^{+5.9}$
&$22.2_{-3.9}^{+4.4}$
&$22.4_{-3.6}^{+4.1}$\\
\hline
\multicolumn{9}{c}{CX14}\\
\hline
\multicolumn{1}{l}{$N_H$}
&\multicolumn{8}{c}{$1.73\pm 0.23$}\\
$kT_\mathit{eff}^{\infty}$
&$67_{-8}^{+6}$
&$66_{-5}^{+4}$
&---
&$60_{-9}^{+6}$
&$66_{-6}^{+4}$
&$64_{-10}^{+7}$
&$66_{-7}^{+5}$
&$65\pm 5$\\
\multicolumn{1}{l}{$\Gamma$}
&\multicolumn{8}{c}{$1.5$}\\
P-l norm.
&$6.9\pm 1.2$
&$2.9\pm 0.7$
&---
&$4.7\pm 1.0$
&$2.0\pm 0.7$
&$7.7\pm 1.9$
&$6.6\pm 1.2$
&$5.4\pm 0.8$\\
Unabs. Thermal Flux
&$5.3_{-3.6}^{+4.6}$
&$5.5_{-2.6}^{+3.2}$
&---
&$3.2_{-2.3}^{+3.3}$
&$5.1_{-2.6}^{+3.0}$
&$4.8_{-3.5}^{+5.0}$
&$5.2_{-3.2}^{+4.2}$
&$5.3_{-2.7}^{+3.4}$\\
Unabs. Flux
&$10.9_{-3.7}^{+4.7}$
&$7.8_{-2.6}^{+3.2}$
&---
&$6.9_{-2.4}^{+3.4}$
&$6.7_{-2.7}^{+3.1}$
&$10.8_{-3.7}^{+5.2}$
&$10.4_{-3.3}^{+4.3}$
&$9.5_{-2.8}^{+3.4}$\\
\hline
\multicolumn{9}{c}{CX15}\\
\hline
\multicolumn{1}{l}{$N_H$}
&\multicolumn{8}{c}{$1.91\pm 0.13$}\\
$kT_\mathit{eff}^{\infty}$
&$77\pm 3$
&$79\pm 3$
&$79\pm 5$
&$65_{-8}^{+5}$
&$74_{-4}^{+3}$
&$69_{-4}^{+5}$
&$80_{-3}^{+2}$
&$71_{-2}^{+3}$\\
\multicolumn{1}{l}{$\Gamma$}
&\multicolumn{8}{c}{1.5}\\
P-l norm.
&$3.5\pm 0.8$
&$3.8\pm 1.0$
&$5.1\pm 2.0$
&$7.1\pm 1.3$
&$3.6\pm 0.8$
&$2.0\pm 1.2$
&$1.8\pm 0.7$
&$1.7\pm 0.5$\\
Unabs. Thermal Flux
&$12.4_{-2.9}^{+3.4}$
&$13.5_{-3.2}^{+3.8}$
&$14.7_{-5.0}^{+6.2}$
&$5.0_{-2.8}^{+3.6}$
&$10.1_{-2.6}^{+3.2}$
&$8.1_{-3.1}^{+3.9}$
&$14.2_{-3.1}^{+3.7}$
&$8.7_{-2.1}^{+2.5}$\\
Unabs. Flux
&$15.1_{-3.0}^{+3.5}$
&$16.6_{-3.3}^{+3.8}$
&$18.6_{-5.2}^{+6.5}$
&$10.7_{-3.0}^{+3.7}$
&$12.9_{-2.7}^{+3.3}$
&$9.6_{-3.2}^{+4.1}$
&$15.6_{-3.2}^{+3.7}$
&$9.9_{-2.1}^{+2.6}$\\
\hline

\end{tabular}
\end{center}
NOTE.---Column density is in units of $10^{22}$ cm$^{-2}$, units of effective temperature are eV, power-law normalization is in units of $10^{-6}$ photons keV$^{-1}$ cm$^{-2}$ s$^{-1}$, and flux is given for the 0.5- 10 keV energy range in units of $10^{-14}$ ergs cm$^{-2}$ s$^{-1}$.  CX9 and CX14 were not detected during observation 3 of Terzan 5 due to the outburst of CX3.
\end{table*}

\section{Discussion}

We have studied the variability of 9 quiescent neutron stars in the globular clusters NGC 6440 and Terzan 5.  In particular, we looked at the thermal variability of these objects.  No objects showed strong evidence for thermal variability between epochs. In all but two cases (CX5 in NGC 6440  when fit using C-statistics and CX1 in NGC 6440 when fit using $\chi^2$ statistics), the effective temperature was found to be constant within the 99\% confidence level.  In the cases in which a possible change in effective temperature is seen, it is marginal and depends on the exact model fitted.   When CX1 in NGC 6440 is fit with $\chi^2$ statistics, it is equally well fit with a variable thermal component or a variable power law.  Therefore, we can neither exclude or require thermal variability in this source.  In CX5 in NGC 6440, it depends on whether we assume a power-law component is present or not. 

Modeling the thermal emission from quiescent neutron stars has been demonstrated to be a promising method for measuring neutron star radii.  Our results support this through showing that a large fraction of sources are not highly variable (see the discussion below).

To put our results into context, we now discuss variability of other quiescent sources.  Of the known neutron star X-ray transients, relatively few sources have been studied over multiple epochs.  The crustal cooling sources, where the neutron stars are seen to cool after the end of an outburst, are one obvious exception \citep[e.g.][]{cackett06}.  This thermal variability is  attributed to cooling, while here we focus on sporadic variability that may indicate accretion is continuing during quiescence.  Of other sources where multiple epochs have been studied, a number of quiescent LMXBs are seen to be steady.  For instance,  \citet{guillot13} studied 4 sources (quiescent LMXBs in M28, NGC 6397, M13 and $\omega$ Cen) over the course of multiple observations and found that they do not exhibit variability.  In fact, for the $\omega$ Cen and NGC 6397 quiescent LMXBs tight constraints of $<2.1\%$ and $<1.4\%$ can be put on their thermal variability \citep{heinke14}, and \citet{servillat12} also discuss the lack of variability in the M28 quiescent LMXB.  Moreover, the spectra of X7 in 47 Tuc \citep{heinke06} were found not to vary.  

On the other hand, there are other quiescent LMXBs that do show variability.   EXO~1745$-$248 in Terzan 5 and SAX~1808.4$-$3658 both show variable hard spectra \citep{degenaar12,heinke09}, but neither has a thermal component.   IGR~J18245$-$2452 in M28 also shows a hard variable spectrum, and shifts between accretion-powered and rotation-powered millisecond pulsations \citep{linares14}.   XMM~J174457$-$28503 was found to show marked variability, though the quiescent spectrum is well-fit by a hard absorbed power-law alone \citep{degenaar14}. However, most quiescent neutron stars require both thermal and non-thermal components to fit their spectra.  Of those that have been observed to be variable, variability in the non-thermal component is usually the cause -- it is rare for a change in the thermal component to be required.  Sources that show this behavior include Aql X-1, SAX~J1750.8$-$2900, Swift~J174805.3$-$244637, IGR~J00291+5934, and we describe these objects in more detail below.

Of these sources, Aql~X-1 has been the best studied, but the cause of its variability cannot be conclusively attributed to a changing column density, power-law component, or effective temperature \citep{rutledge01,campana03,cackett11,cotizelati14}.  SAX~J1750.8$-$2900 was observed over the course of five weeks in early 2013.  The source increased in luminosity by a factor of 3 -- 4 for a period between 5 -- 16 days, and then decayed to its originally observed luminosity within a week.  \citet{wijnands13} compared the quiescent spectra to that previously fit by \citet{lowell12} from observations two years prior, and found the parameters of thermal models not to vary.  The flare spectrum required a power-law component.  When fit with a combined thermal and power-law model, the effective temperature was not constrained tightly enough for comparison, though the data were well-fit \citep{wijnands13}.   \citet{bahramian14} found the flux of Swift~J174805.3$-$244637 (the third X-ray transient in Terzan 5) to vary in the power-law component only.  

\citet{jonker05} found a change in the effective temperature of IGR~J00291+5934 when fitting a single-component model, but the statistics do not allow for a well-constrained fit of a combined absorbed neutron star atmosphere and power-law model.  The conclusions are therefore limited, and do not make a strong case for thermal variability.  

While the thermal component of the crustal cooling sources generally shows a smooth cooling trend, several show sporadic variability. For instance,  EXO~0748$-$676 has a variable power-law component in the quiescent spectrum, though this does not appear to affect the thermal component, which shows a smooth decay \citep{degenaar14b}.  Another example is XTE~J1701$-$462 which also has a variable power-law component \citep{fridriksson10,fridriksson11}.  However, it also shows more dramatic short-term flares (e-folding time of approximately 3 days) at a flux of about 20 times the cooling trend \citep{fridriksson10,fridriksson11}.  During these short flares the measured temperature is significantly higher, but, the temperature returns to the cooling trend quickly afterwards.

Thermal variability in the quiescent spectrum is also seen in the nearby, bright neutron star LMXB Cen X-4. Its highly variable spectra cannot be fit without a changing thermal component \citep{cackett10, cackett13_cenx4, bernardini13}.   In summary, while the 11 quiescent sources discussed above have shown quiescent variability, of the 7 which have a thermal component, only two of them show significant thermal variability not associated with crustal cooling (Cen X-4 and XTE~J1701$-$462).

Of the additional 9 sources we examined for variability in this paper, 7 of them do not show strong evidence for variability, including all of the objects in Terzan 5.  A previous study of CX1 in NGC 6440 showed variability attributable to the power-law component between the first two epochs \citep{cackett05}.  The additional observations studied here maintain this conclusion when fit with the C-statistic.  When fit with $\chi^2$ statistics, only 3 observations could be analyzed.  The spectra were equally well fit by a variable thermal component or a variable power law.  Therefore, the results are inconclusive and we can neither require or exclude thermal variability in this source. CX5 in NGC 6440 could only be fit with C-statistics, due to the low number of counts in all observations.  It was found to be variable when fit without a power-law component, but, when a power-law component is included the significance of the variability drops, hence, the low signal-to-noise ratio prevents us from making strong conclusions about the variability in this source.  

We also consider whether the variability seen in Cen~X-4 would be detected in our much lower signal-to-noise ratio data here.  By the same method we used earlier for the sources studied here, we used results from \citet{cackett10,cackett13_cenx4} to measure the level of thermal variability.  Cen X-4 has a maximum 1$\sigma$ deviation of 13 eV from the mean of 58 eV, a deviation of 22\%.  XTE~J1701$-$462 has even larger thermal variability, with the flare approximately 25 eV hotter than surrounding points. Thus, though the LMXBs studied in this paper are faint sources (and hence the constraints on temperature are not as tight as for Cen X-4) if they showed the level of variability seen in Cen X-4, it would be large enough to be significant here. 

Therefore, to the 5 known stable sources, and 7 known variable sources that have a thermal component, we add 7 non-variable sources and 1 source that varies, and an another source that we cannot conclusively say is variable or not.  Therefore, only 2 sources show strong evidence of thermal variability amongst the 21 objects with a quiescent thermal component discussed in this paper, showing that the vast majority of quiescent sources are not thermally variable within current measurement constraints.  However, as discussed in \citet{degenaar14} there may be an additional group of sources that may not be in what is typically considered as a truly quiescent state.  \citet{degenaar14}  denote this as an intermediate state (between quiescence and outburst), citing KS~1741$-$293, XTE~J1701$-$462, and GRS~1741$-$2853 as examples of sources during which luminosity increases for a short period of time, along with SAX~1750.8$-$2900, GRO~1744$-$28, 4U~1608$-$52, SAX~J1747.0$-$2853 and XMM~J174457$-$2850.3 as examples of sources with a slighter, longer increase in luminosity.  The majority of these sources have not been study for spectral variability and thus those sources are not included in our count of spectrally variable sources.  

 \citet{wijnands14} reviewed literature for spectral fits of NS LMXBs over the 0.5$-$10 keV range fit with only an absorbed power-law with intermediate luminosities (between $10^{34}$ and $10^{36}$ ergs s$^{-1}$), a factor of at least 10 greater than the sources in our study.  They found that the photon index and X-ray luminosity are inversely related and suggest that accretion may contribute to thermal components as well in this luminosity range.  While the sources we study here are below this luminosity range, it is reasonable that as we begin to increase studies of sources of intermediate luminosities, our understanding of the physical mechanisms distinguishing quiescent neutron star binaries from those in outburst, as well as links between them will begin to be better understood.  \citet{wijnands14} conclude that residual accretion causes both the thermal and non-thermal components in the intermediate luminosity sources they study.  Furthermore, they suggest the same mechanisms may be present below $10^{34}$ erg s$^{-1}$ in the types of sources we study here.  Although we see little thermal variability in the majority of the sources we cannot rule out such a scenario.  It may be at these lower luminosities the thermal radiation from the cooling neutron star dominates over any additional thermal radiation caused by low level accretion.

It is widely speculated that residual accretion is the cause of the observed quiescent variability \citep[e.g][]{cackett10, cackett11, cackett13_cenx4, bernardini13}.  It is interesting to consider whether any binary parameters (such as orbital period), which could determine how accretion occurs in quiescence, correlate with quiescent variability.  Unfortunately, not all sources have known orbital periods.  Of the stable sources, the quiescent LMXBs in $\omega$ Cen and NGC 6397 both have implied short orbital period \citep[less than 1.5 hours][]{haggard04, heinke14}.  While, the tendency is for the variable objects to have orbital periods longer than this:  IGR J18245-2452 in M28 has an orbital period of 11 hours \citep{papitto13}, Aql~X-1 has a period of 19 hours \citep{chevalier91}, Cen X-4 has an orbital period of 15.1 hours \citep{chevalier89} and CX1 in NGC 6440 has an orbital period of 8.7 hours \citep{altamirano08}.  However, there are clear exceptions to this, for example variable sources IGR J00291+5934 and SAX~1808.4$-$3658 also have short orbital periods  \citep[2.46 hours and 2.01 hours, respectively,][]{tomsick05,chakrabarty98}.  Thus, there is not strong evidence yet for a correlation between quiescent variability and orbital period, and further investigation to the cause of variability is needed.

\begin{table*}
\small
\caption{Spectral parameters for CX1 in NGC 6440 using $\chi^2$ statistics}
\label{tab:cx1bah}
\begin{center}
\renewcommand{\arraystretch}{1.2}
\renewcommand{\tabcolsep}{0.25cm}
\begin{tabular}{lccc}
\hline
\hline
\multicolumn{1}{l}{Model Parameter}
&\multicolumn{1}{c}{Obs 1}
&\multicolumn{1}{c}{Obs 2}
&\multicolumn{1}{c}{Obs 3}\\
\hline
\multicolumn{4}{c}{Power-law normalization free}\\
\hline
\multicolumn{1}{l}{$N_H$}
&\multicolumn{3}{c}{$0.76\pm0.05$}\\
$kT_\mathit{eff}^{\infty}$
&\multicolumn{3}{c}{$90\pm2$}\\
\multicolumn{1}{l}{$\Gamma$}
&\multicolumn{3}{c}{$3.1\pm0.5$}\\
P-l norm. $(10^{-6})$
&$40.8_{-12.4}^{+16.5}$
&$\le 2.9$
&$\le 11.2$\\
Unabs. Thermal Flux (0.5- 10 keV) $(10^{-14}$ ergs cm$^{-2}$ s$^{-1}$)
&\multicolumn{3}{c}{$11.5\pm0.6$}\\
Unabs. Flux (0.5- 10 keV) $(10^{-14}$ ergs cm$^{-2}$ s$^{-1}$)
&$23.1\pm 1.9$
&$13.4\pm0.7$
&$19.3\pm0.9$\\
$\mathrm{F}_\mathrm{var}$
&\multicolumn{3}{c}{$0.25\pm 0.04$}\\
\multicolumn{1}{l}{$\chi^{2}$}
&\multicolumn{3}{c}{34.42 for 34 d.o.f.}\\
\multicolumn{1}{l}{Reduced $\chi^{2}$}
&\multicolumn{3}{c}{1.012}\\
\hline
\multicolumn{4}{c}{Effective temperature free}\\
\hline
\multicolumn{1}{l}{$N_H$}
&\multicolumn{3}{c}{$0.76\pm0.04$}\\
$kT_\mathit{eff}^{\infty}$
&$101_{-3}^{+2}$
&$89_{-3}^{+1}$
&$92_{-1}^{+2}$\\
$\Gamma$&\multicolumn{3}{c}{1.5}\\
P-l norm. $(10^{-6})$&\multicolumn{3}{c}{$\le0.7$}\\
Unabs. Thermal Flux (0.5- 10 keV) $(10^{-14}$ ergs cm$^{-2}$ s$^{-1}$)
&$18.7\pm1.3$
&$10.1\pm1.0$
&$12.4\pm0.8$\\
Thermal $\mathrm{F}_\mathrm{var}$
&\multicolumn{3}{c}{$0.32\pm0.04$}\\
Unabs. Flux (0.5- 10 keV) $(10^{-14}$ ergs cm$^{-2}$ s$^{-1}$)
&$18.7\pm1.3$
&$10.1\pm1.0$
&$12.4\pm0.8$\\
$\mathrm{F}_\mathrm{var}$
&\multicolumn{3}{c}{$0.32\pm0.04$}\\
\multicolumn{1}{l}{$\chi^{2}$}
&\multicolumn{3}{c}{37.71 for 37 d.o.f.}\\
\multicolumn{1}{l}{Reduced $\chi^{2}$}
&\multicolumn{3}{c}{1.019}\\
\hline

\end{tabular}
\end{center}
NOTE.---A mass of 1.4 M$_{\odot}$ and radius of 10 km was assumed for the \texttt{nsatmos} model.  
\end{table*}

\begin{table*}
\begin{center}
\tiny

\caption{Spectral parameters for CX9 and CX12 in Terzan 5 using $\chi^2$ statistics}
\label{tab:t5bah}
\renewcommand{\arraystretch}{1.2}
\renewcommand{\tabcolsep}{0.15cm}
\begin{tabular}{lcccccccc}
\hline
Model Parameter & Obs. 1 & Obs. 2 & Obs 3 & Obs 4 & Obs 5 & Obs 6 & Obs 7 & Obs 8 \\
\hline
\multicolumn{9}{c}{CX9}\\
\hline
\multicolumn{9}{c}{All parameters tied}\\
\hline
$N_{\rm H}$ & \multicolumn{8}{c}{$1.85\pm0.13$}\\
$kT_\mathit{eff}^{\infty}$ &\multicolumn{8}{c}{$93_{-6}^{+4}$}\\
$\Gamma$ &\multicolumn{8}{c}{$1.4\pm1.0$}\\
P-L norm. &\multicolumn{8}{c}{$4.5_{-3.3}^{+11.0}$}\\
Unabs. Thermal Flux &\multicolumn{8}{c}{$12.9\pm0.9$}\\
Unabs. Flux &\multicolumn{8}{c}{$16.7\pm1.0$}\\
$\chi^{2}$ &\multicolumn{8}{c}{46.17 for 35 d.o.f.}\\
Reduced $\chi^{2}$ &\multicolumn{8}{c}{1.319}\\
\hline
\multicolumn{9}{c}{Power-law normalization free}\\
\hline
$N_{\rm H}$ &\multicolumn{8}{c}{$1.90\pm0.12$}\\
$kT_\mathit{eff}^{\infty}$ & \multicolumn{8}{c}{$96_{-3}^{+2}$}\\
$\Gamma$ &\multicolumn{8}{c}{$0.5_{-1.0}^{+0.8}$}\\
P-l norm. &$2.4_{-1.7}^{+3.6}$ & $0.9_{-0.8}^{+2.1}$ &--- & $1.0_{-0.8}^{+2.3}$ &$1.0_{-0.7}^{+1.9}$ &--- &$2.7_{-1.9}^{+4.4}$ &$1.1_{-0.8}^{+2.2}$\\
Unabs. Thermal Flux &\multicolumn{8}{c}{$18.6_{-9.5}^{+25.7}$}\\
Unabs. Flux & $27.9_{-9.8}^{+25.9}$ & $22.4_{-9.9}^{+25.9}$ &--- & $22.5_{-9.6}^{+25.8}$ &$22.3_{-9.6}^{+25.8}$ &--- &$28.7_{-9.7}^{+25.8}$ &$22.6_{-9.6}^{+25.8}$\\
$\mathrm{F}_\mathrm{var}$ & \multicolumn{8}{c}{---}\\
$\chi^{2}$ & \multicolumn{8}{c}{32.40 for 30 d.o.f.}\\
Reduced $\chi^{2}$ &\multicolumn{8}{c}{1.080}\\
\hline
\multicolumn{9}{c}{Effective temperature free}\\
\hline
$N_{\rm H}$ &\multicolumn{8}{c}{$1.90\pm0.13$}\\
$kT_\mathit{eff}^{\infty}$ &$95_{-5}^{+4}$ &$93_{-6}^{+4}$ &--- &$96_{-5}^{+4}$ &$92_{-6}^{+3}$ &--- &$95_{-5}^{+4}$ &$95_{-5}^{+4}$\\
$\Gamma$ &\multicolumn{8}{c}{$1.3\pm1.0$}\\
P-l norm.  &\multicolumn{8}{c}{$3.6_{-2.7}^{+8.8}$}\\
Unabs. Thermal Flux &$14.1\pm1.8$ &$12.9\pm2.0$ &--- &$15.1\pm2.2$ &$12.1\pm1.9$ &--- &$13.9\pm2.0$ &$14.5\pm1.9$\\
Thermal $\mathrm{F}_\mathrm{var}$ &\multicolumn{8}{c}{---}\\
Unabs. Flux &\multicolumn{8}{c}{$3.8\pm0.5$}\\
$\chi^{2}$ &\multicolumn{8}{c}{42.95 for 30 d.o.f.}\\
Reduced $\chi^{2}$ &\multicolumn{8}{c}{1.432}\\
\hline
\multicolumn{9}{c}{CX12}\\
\hline
\multicolumn{9}{c}{All parameters tied}\\
\hline
$N_{\rm H}$ &\multicolumn{8}{c}{$1.89\pm0.17$}\\
$kT_\mathit{eff}^{\infty}$ & \multicolumn{8}{c}{$86_{-15}^{+7}$}\\
$\Gamma$ &\multicolumn{8}{c}{$2.5\pm0.7$}\\
P-l norm. &\multicolumn{8}{c}{$17.3_{-11.3}^{+25.2}$}\\
Unabs. Thermal Flux &\multicolumn{8}{c}{$8.9\pm1.1$}\\
Unabs. Flux & \multicolumn{8}{c}{$15.2\pm1.2$}\\
$\chi^{2}$ &\multicolumn{8}{c}{48.70 for 34 d.o.f.}\\
Reduced $\chi^{2}$ &\multicolumn{8}{c}{1.432}\\
\hline
\multicolumn{9}{c}{Power-law normalization free}\\
\hline
$N_{\rm H}$ &\multicolumn{8}{c}{$1.72\pm0.19$}\\
$kT_\mathit{eff}^{\infty}$ &\multicolumn{8}{c}{$\le85$}\\
$\Gamma$ &\multicolumn{8}{c}{$3.1\pm0.5$}\\
P-l norm. &$32.7_{-22.1}^{+18.3}$ &$44.2_{-26.9}^{+24.3}$ &--- &$42.0_{-23.7}^{+25.0}$ &$28.8_{-17.5}^{+19.6}$ &--- &$46.2_{-14.7}^{+28.3}$ &$46.4_{-26.9}^{+13.2}$\\
Unabs. Thermal Flux
&\multicolumn{8}{c}{$2.7\pm1.5$}\\
Unabs. Flux &$13.0\pm2.0$ &$16.5\pm2.2$ &--- &$15.8\pm2.4$ &$11.7\pm2.2$ &--- &$17.1\pm 2.3$ &$17.2\pm2.1$\\
$\mathrm{F}_\mathrm{var}$ &\multicolumn{8}{c}{$\le0.19$}\\
$\chi^{2}$ &\multicolumn{8}{c}{36.09 for 29 d.o.f.}\\
Reduced $\chi^{2}$  &\multicolumn{8}{c}{1.245}\\
\hline
\multicolumn{9}{c}{Effective temperature free}\\
\hline
$N_{\rm H}$ &\multicolumn{8}{c}{$1.73\pm0.13$}\\
$kT_\mathit{eff}^{\infty}$ &$76_{-7}^{+12}$ &$83_{-6}^{+9}$ &--- &$77_{-8}^{+12}$ &$\le80$ &--- &$84_{-5}^{+10}$ &$84_{-5}^{+9}$\\
$\Gamma$ &\multicolumn{8}{c}{$2.8\pm0.5$}\\
P-l norm. &\multicolumn{8}{c}{$27.5_{-1.0}^{+10.0}$}\\
Unabs. Thermal Flux &$5.2\pm2.1$ &$7.4\pm2.1$ &--- &$5.3\pm2.4$ &$\le1.8$ &--- &$8.0\pm2.2$ &$8.0\pm2.0$\\
Thermal $\mathrm{F}_\mathrm{var}$ &\multicolumn{8}{c}{$0.38\pm0.17$}\\
Unabs. Flux 
&$13.8\pm 2.2$
&$16.1\pm2.3$
&---
&$13.9\pm2.5$
&$8.9_{-0.8}^{+1.7}$
&---
&$16.6\pm2.3$
&$16.6\pm2.1$\\
$\mathrm{F}_\mathrm{var}$ &\multicolumn{8}{c}{$0.38\pm0.17$}\\
$\chi^{2}$ &\multicolumn{8}{c}{30.46 for 29 d.o.f.}\\
Reduced $\chi^{2}$ &\multicolumn{8}{c}{1.050}\\
\hline
\end{tabular}
\end{center}
NOTE.---Column density is in units of $10^{22}$ cm$^{-2}$, units of effective temperature are eV, power-law normalization is in units of $10^{-6}$ photons keV$^{-1}$ cm$^{-2}$ s$^{-1}$, and flux is given for the 0.5- 10 keV energy range in units of $10^{-14}$ ergs cm$^{-2}$ s$^{-1}$.  CX9 was not detected during observation 3 of Terzan 5 due to the outburst of CX3.  Observation 3 of CX12 and observation 6 of CX9 and CX12 had too low a photon count to analyze using $\chi^2$ statistics.  
\end{table*}

\section*{Acknowledgements}
A. R. W. gratefully acknowledges the National Science Foundation for support through a Research Experience for Undergraduates program at Wayne State University (NSF Grant No. PHY-1156651).  The scientific results reported in this article are based on data obtained from the Chandra Data Archive.

\bibliographystyle{mn2e}
\bibliography{qNS}

\end{document}